\documentclass[12pt]{iopart}
\usepackage{multirow}
\usepackage{graphicx}
\usepackage{iopams}
\usepackage{bm}
\usepackage{color}
\newcommand\cc{\multicolumn{1}{|c|}}

\begin{document}

\title[Plane-layer Rayleigh-B\'enard convection up to $Ra=10^{11}$]{Plane-layer Rayleigh-B\'enard convection up to $Ra=10^{11}$: Near-wall fluctuations and role of initial conditions}

\author{Roshan J. Samuel and J\"org Schumacher}
\address{Institute of Thermodynamics and Fluid Mechanics, Technische Universit\"{a}t Ilmenau, D-98684 Ilmenau, Germany}
\ead{joerg.schumacher@tu-ilmenau.de}
\vspace{10pt}

\begin{abstract}
We study turbulent Rayleigh-B\'enard convection through direct numerical simulations in a three-dimensional plane layer of aspect ratio 4 for Rayleigh numbers $10^5 \leq Ra \leq 10^{11}$ and Prandtl number $Pr=0.7$. We summarize the height-dependent statistics of velocity and temperature fluctuations and corresponding scalings with the Rayleigh number. We include an analysis on the role of coherent and incoherent flow regions near the wall for global heat transfer. Furthermore, we investigate the dependence of turbulent transport on a finite-amplitude sinusoidal shear flow added at time $t=0$, which either freely decays in a long transient or remains existent when a steady sinusoidal volume forcing is added. In the latter case, weak logarithmic near-wall layers are formed, however, with von K\'arm\'an and offset constants that differ from standard values. The typical magnitude of both coefficients, and thus a full turbulent boundary layer of velocity and temperature, is re-established only for a switch from sinusoidal to constant pressure gradient driving of the flow. In all cases, except for the constant pressure gradient-driven flow, no enhancement of global turbulent heat and momentum transfer within error bars is detected, even though the sinusoidal amplitude is of the order of the characteristic free-fall velocity.
\end{abstract}

\vspace{2pc}

\submitto{\FDR}

\section{Introduction}
\label{sec:intro}

Rayleigh-B\'enard convection (RBC) is a benchmark problem of fluid turbulence, that has withstood the test of time to remain a fruitful field of research. Spurred by its relevance to atmospheric flows, industrial cooling and heat exchange systems, planetary boundary layers and a host of other applications, RBC continues to be a subject of both experimental and numerical studies~\cite{Niemela:Nature2000,Ahlers:RMP2009,Chilla2012,Lohse2024}. The canonical form of RBC invokes a plane layer of fluid confined between two parallel, infinitely extended horizontal plates. Unstable stratification requires a colder top and hotter bottom plate~\cite{Verma:BDF}. Under the influence of gravity which acts along the vertical direction, a buoyancy-driven convective flow arises, transferring the heat between the plates, on average from the bottom to the top. The driving coherent flow structures for this heat transfer are the thermal plumes, whose structure and impact on flow properties has been a subject of considerable scrutiny~\cite{Baburaj:JFM2005,Zhou:PRL2007,Schumacher:PRE2016}. These plumes originate from thermal instabilities in the boundary regions near the walls~\cite{Malkus1954,Howard1966} for sufficiently high temperature gradients. Understanding their role for turbulent heat and momentum transfer requires a height-dependent investigation of the fluctuation statistics of both temperature and velocity, which is one focus of the present work. A similar study for confined convection in a closed cylinder was also performed by~\cite{Emran:JFM2008}.

Although thermal plumes drive the convective turbulence, the dynamics is not purely plume-dominated. In \cite{Samuel2024}, it was shown that the viscous near-wall region is a patchwork of local shear-dominated coherent and plume-detaching incoherent regions in a nearly constant percentage area ratio of approximately 40:60, which is in line with the absence of a global unidirectional mean flow in this plane-layer RBC configuration. The near-wall layers of temperature and, in particular, of velocity  are dominated by fluctuations for the whole range of Rayleigh numbers between $10^5$ and $10^{11}$. In \cite{Shevkar2025}, it was demonstrated that thermal plumes are organized in a hierarchical and self-similar network that gets coarser with increasing wall distance. The network dynamics can be described as a continued coagulation and reformation process. A basic building block of the dynamics are local marginally unstable plumes --- segments of the thermal boundary layer, which has been studied in several models \cite{Howard1966,Theerthan1998,Waleffe2015}. This self-organization of the near-wall flow in extended plane layers seems to be imperative for the efficient transport of heat that follows closely a classical 1/3 scaling with respect to Rayleigh number; it leads to near-wall layers that are different from standard boundary layers in wall-bounded shear flows.

The dichotomy of coherent (shear dominance or plume impact) and incoherent (plume detachment) regions, particularly their role for the heat transfer, was also investigated for two-dimensional RBC in \cite{Zhu:PRL2018} and \cite{He2024}. Furthermore, confined convection in closed cells at an aspect ratio close to 1 or smaller is dominated by a large-scale circulation (LSC) \cite{Ahlers:RMP2009}. A dynamically analogous description of the boundary region flow into plume-detachment and post-plume phases also shows significant differences between both regimes~\cite{Shi:JFM2012}. In the present work, we will explore the strategy of isolating these two sides of the boundary layer flow by thresholding the wall-shear stress field in order to compare the findings with those of \cite{Samuel2024}. One of our goals in disentangling the coherent and incoherent regions of the boundary layer in different ways is to quantify differences in heat transfer between these two regions. This is a second motivation point for the present study.

It has been hypothesized that the turbulent flow passes through a transition at very high Rayleigh number $Ra$ (which quantifies the thermal driving in RBC),  which leads to a heat transfer scaling law $Nu\sim Ra^{\gamma}$ with an exponent larger than the classical one, $\gamma>1/3$. The Nusselt number $Nu$ is a dimensionless measure of turbulent heat transfer. This regime of RBC is termed the ``ultimate regime''~\cite{Kraichnan:PF1962,Spiegel:ApJ1963}. Significant efforts have been expended towards attaining very high Rayleigh numbers in turbulent convection in laboratory experiments~\cite{Niemela:Nature2000,Chavanne:PRL1997,Ahlers:NJP2011,Urban2011,He2012} and three-dimensional numerical simulations \cite{Stevens2011,Iyer2020,Samuel2024,Tiwari2025} to detect this regime. See also \cite{Niemela:JFM2003,Skrbek:JFM2015,Doering:PNAS2020,Roche2020,Lindborg2023,Lohse2024} for different conclusions on its existence. A more recent study attempts to explain the absence of this transition by showing that the positive and negative contributions to convective heat flux are nearly equal in convection confined by horizontal plates above and below \cite{Tiwari:PNAS2025}.

In \cite{Roche2020} and later in \cite{Shishkina2023,Lohse2024}, it was proposed that such a regime change, if existing, is connected to a transition to turbulence of the near-wall layers --- analogous to a subcritical transition to turbulence in wall-bounded shear flows \cite{Avila2023}. This would imply the coexistence of two global flow regimes (or states) in turbulent RBC \cite{Shishkina2023}, similar to the laminar and turbulent state in wall-bounded shear flows with the latter state trapped in a chaotic repeller that is formed by a skeleton of linearly unstable exact coherent states \cite{Hof2004,Eckhardt2007}. To carry such a loose analogy even further: a coexistence would be in line with a sensitive dependence on the specific form of finite-amplitude perturbations that trigger hysteretic transitions between both macrostates; see for example \cite{Moehlis2004,Hof2006} for wall-bounded shear flows.

A resulting further motivation for the present work is thus to investigate the dependence of the RBC on the initial condition. Do different initial conditions kick the RBC system into different turbulent macrostates (or attractors) with different near-wall dynamics? If yes, how do they affect the global transport? DNS of RBC typically start with a linear temperature profile across the layer, the diffusive and quiescent equilibrium state. An infinitesimal perturbation brings RBC into a fully turbulent state once $Ra$ is high enough. In the present work, we address the two questions from above in a series of DNS at Rayleigh number $Ra=10^9$. The susceptibility to relax into different RBC macrostates is probed by different initial perturbations of the diffusive equilibrium. To this end, we add a sinusoidal shear mode $u_x(z, t=0)=\tilde{A}_i \sin(\pi z)$ in agreement with incompressibility and boundary conditions and a {\em finite amplitude} $\tilde{A}_i$ to compare the long-term evolution of initially different RBC runs with respect to each other. We find that all runs with different initial conditions relax to the same turbulent attractor and not into different turbulent macrostates. This is probed by global transport measures, Nusselt and Reynolds number, of these runs.

This idea is then further extended by adding a steady volume forcing $f_x$ to the momentum balance that maintains this sinusoidal shear mode $u_x(z)$, see e.g. \cite{Schumacher2001}. Even in this case, the Nusselt and Reynolds number, the global measures of heat and momentum transfer, remains unchanged within the error bars compared to the standard case with $u_x(z,t=0)=0$ and $f_x=0$. Interestingly, for the run with $f_x\ne 0$, we detect a logarithmic region in the mean temperature and velocity profiles. However, their von K\'arm\'an and offset constants differ from previous mixed convection studies in pressure-driven plane Poiseuille \cite{Pirozzoli:JFM2017,Hamman2018} or plane Couette geometries \cite{Blass:JFM2020,Blass:JFM2021}, both with heated and smooth walls. Furthermore, for this non-standard logarithmic near-wall layer, formed in the case of a steady but moderate volume forcing, we show that the global transport properties remain unaffected within the error bar. We conclude that this shear forcing is still too weak; the flow, even though perturbed by a finite amplitude, remains dominated by buoyancy forces. Once we substitute sinusoidal with a constant pressure gradient forcing, a fully developed standard turbulent boundary layer is formed for both fields with coefficients as in shear flows.

The outline of this paper is as follows. In \sref{sec:numerics}, we present the numerical model. In \sref{sec:hds}, we summarize the height-dependent statistics of temperature and velocity, which complements our earlier study in \cite{Samuel2024}. \Sref{sec:decompose} summarizes the refined analysis with respect to coherent and incoherent regions. \Sref{sec:shear} discusses the impact of different shear-flows on convective turbulence. This discussion includes decaying sinusoidal modes of different amplitudes, a steadily forced sinusoidal mode, and a steady pressure gradient. We finally summarize and conclude our findings in \sref{sec:conclusions}.

\section{Governing equations and numerical method}
\label{sec:numerics}

We solve the incompressible Navier-Stokes equations with the Boussinesq approximation coupling the velocity and temperature fields, ${\bm u}({\bm x}, t)$ and $T({\bm x}, t)$ respectively. These partial differential equations are solved in the non-dimensional form,
\begin{eqnarray}
\frac{\partial {\bm u}}{\partial t} + ({\bm u} \cdot {\bm \nabla}) {\bm u} = -{\bm \nabla}p + T \hat{{\bm z}} + \sqrt{\frac{Pr}{Ra}} \, \nabla^2 {\bm u},
\label{eq:u} \\
\frac{\partial T}{\partial t} + ({\bm u} \cdot {\bm \nabla}) T = \frac{1}{\sqrt{PrRa}} \, \nabla^2 T,
\label{eq:T} \\
{\bm \nabla} \cdot {\bm u} = 0.
\label{eq:m}
\end{eqnarray}
The two dimensionless parameters are Rayleigh number $Ra$ and Prandtl number $Pr$, which are defined as
\begin{equation}
Ra=\frac{g\alpha \Delta T H^3}{\nu \kappa} \quad \mathrm{and} \quad Pr=\frac{\nu}{\kappa}.
\label{eq:RaPr}
\end{equation}
Here, $g$ denotes the acceleration due to gravity, $\alpha$ is the thermal expansion coefficient, $\Delta T$ is the temperature difference between the hot and cold plates, $H$ is the vertical separation between the plates, $\nu$ is the kinematic viscosity, and $\kappa$ is the thermal diffusivity. These dimensional quantities are also used to define the free-fall velocity, $U_f = \sqrt{\alpha g \Delta T H}$, as the characteristic unit of the velocity field. Correspondingly, characteristic length, time and temperature scales are $H$, $H/U_f$, and $\Delta T$ respectively. 

The RBC flow reacts with a turbulent heat and momentum transfer to the set of input parameters, $\{Ra, Pr\}$. The corresponding dimensionless numbers are the Nusselt number $Nu$ for the turbulent heat and the Reynolds number $Re$ for the turbulent momentum transfer. They are given by (all quantities are dimensionless)
\begin{equation}
Nu=-\frac{\partial \langle T\rangle_{A,t}}{\partial z}\Bigg|_{z=0} \quad \mbox{and} \quad Re=\sqrt{\frac{Ra}{Pr}} U^v_{\rm rms}\,,
\label{eq:NuRe}
\end{equation}
where $\langle\cdot\rangle_{A,t}$ is a combined average over cross-section $A$ and simulation time $t$ in statistically steady state. The root mean square velocity $U^v_{\rm rms}=\langle {\bm u}^2\rangle^{1/2}_{V,t}$ with a combined volume-time average $\langle\cdot\rangle_{V,t}$. The Boussinesq equations are solved in a Cartesian box of size $L \times L \times H$, where $L$ denotes both the length and width of the box. We choose a domain of aspect ratio $\Gamma = L/H = 4$ for our simulations, so that the cross-sectional area $A = 16$ in dimensionless units. Together with the periodic boundary conditions on the sides, this setup aims to approximate an infinitely extended plane layer of fluid confined between the top and bottom no-slip walls, which are isothermally maintained at $T$ and $T + \Delta T$ respectively. Further details on this choice of parameters are presented in \cite{Samuel2024}. 

The governing equations are solved using spectral element method (SEM), where the computational domain is discretized into finite elements, on each of which the fields are approximated by Lagrangian interpolation basis functions in all three space dimensions, offering an accuracy of spectral methods~\cite{Fischer1997}. We use the GPU-accelerated SEM solver NekRS~\cite{Fischer2022}, which is a descendant of the Nek5000 code. Our simulations span a range of $Ra$ from $10^5$ to $10^{11}$ at a fixed Prandtl number of $Pr=0.7$. Of particular focus is the case at $Ra=10^9$, with which we present additional studies on the effect of initial condition and forced shear flow in \sref{sec:shear}. The simulations are performed with sufficient spectral resolution, and the planar averaged profiles are also well converged. This is described and discussed in detail in ~\cite{Samuel2024}. There, we also showed that the thermal boundary layers are well-resolved, with at least 16 spectral collocation points within them for all the cases.

\section{Height-dependent statistics of velocity and temperature}
\label{sec:hds}

\begin{figure}
  \centerline{\includegraphics[width=1.0\textwidth]{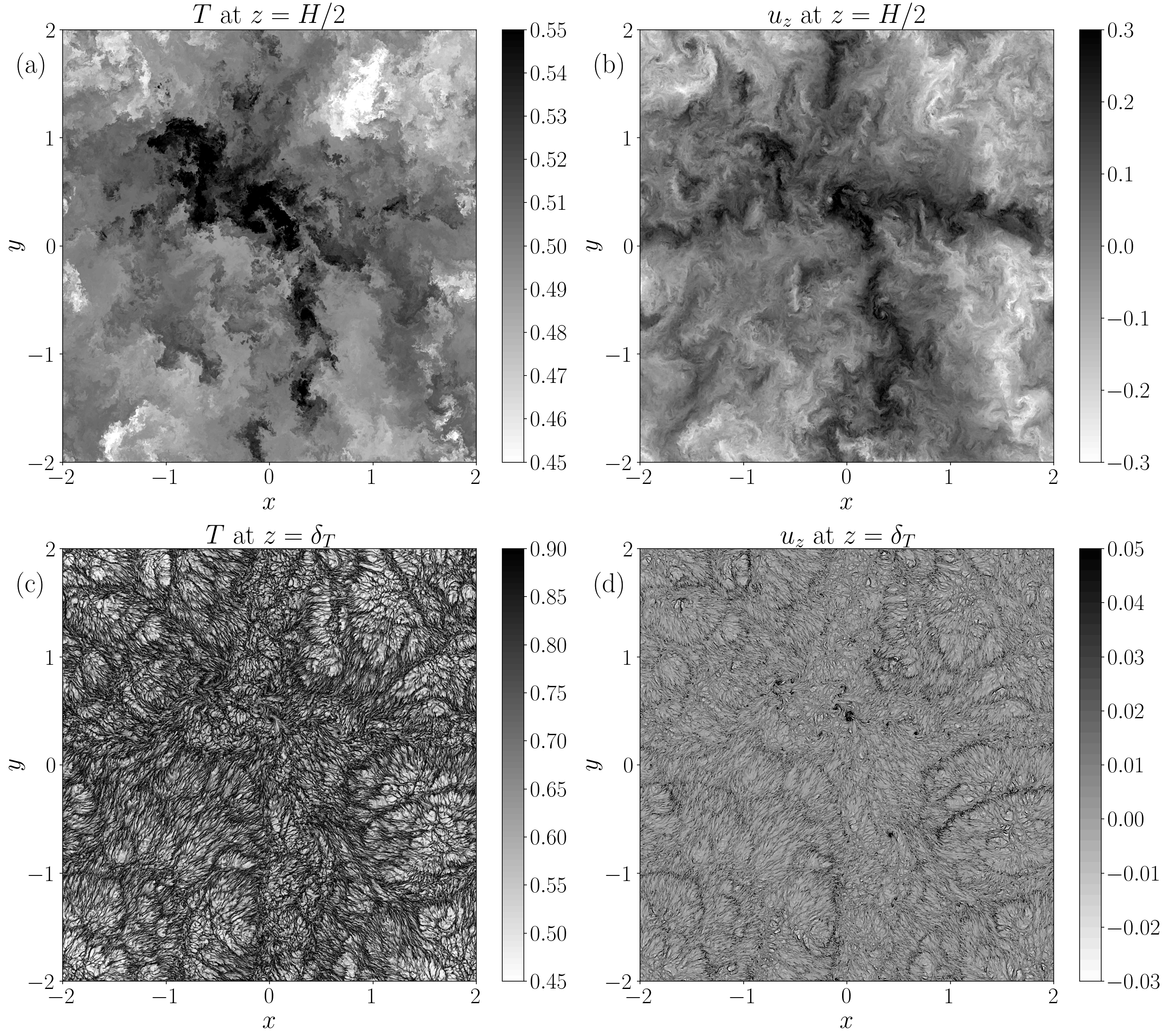}}
  \caption{Contours of temperature (a, c) and vertical component of velocity (b, d) on a horizontal cross-section ($x$-$y$ plane) from a snapshot at $Ra=10^{11}$. The upper and lower rows shows the fields at $z=H/2$ and $\delta_T$ respectively. The structure of the flow is more sharply delineated at the thermal boundary layer where plumes dominate the flow. At the midplane in the bulk, the fields are thoroughly turbulent and well-mixed. Appropriate colorbar cut-offs have been chosen for increased contrast and clear visibility of structures.}
\label{fig:t_uz_field}
\end{figure}

The structure of convective flow is markedly different in the bulk and the boundary layer, especially at higher Rayleigh numbers where the boundary layers become significantly thinner.  As a result, a systematic study of thermal convection requires one to disentangle the two regions and their contributions to the overall heat and momentum transfer~\cite{Ahlers:RMP2009,Emran:JFM2008}. We define the thermal boundary layer thickness $\delta_T$ at both walls as the distances of the local maxima of temperature fluctuations from their respective walls, and this value is practically same as $1/(2Nu)$ which we detailed in \cite{Samuel2024}.

Figure~\ref{fig:t_uz_field} demonstrates the distinction between bulk and boundary layer (BL) flow using the temperature $T$ and vertical velocity component $u_z$ for a snapshot at $Ra=10^{11}$. A well-defined and fine-grained plume structure is visible in the thermal boundary layer at $z=\delta_T$ in the lower row, whereas in the bulk at $z=H/2$ (upper row), the temperature field is well-mixed and the velocity field is turbulent. Nevertheless, we also note that there is no sharp transition from bulk to boundary layer. Rather, there is a self-similar hierarchy of plumes which successively aggregate as we transition from boundary layer to bulk, see \cite{Shevkar2025} for this analysis. As a result, it is important to investigate the fluctuation intensities and statistics of velocity and temperature fields at different heights. In the next subsections, we choose heights of the analysis planes that are multiples of the boundary layer thickness, $\delta$, to obtain the scaling of fluctuation intensities with $Ra$ and probability distribution functions. This thickness $\delta$ of the near-wall layer can be determined even in absence of a global mean flow. We will thus use two thickness scales, one based on the temperature, $\delta_{T,{\rm rms}}$, and one based velocity fluctuation profiles, $\delta_{U,{\rm rms}}$. The corresponding thickness is given by the distance to the first local maximum to the wall (see \sref{sec:profiles}).

\subsection{Vertical profiles of velocity and temperature fluctuations}
\label{sec:profiles}

\begin{figure}
  \centerline{\includegraphics[width=0.8\textwidth]{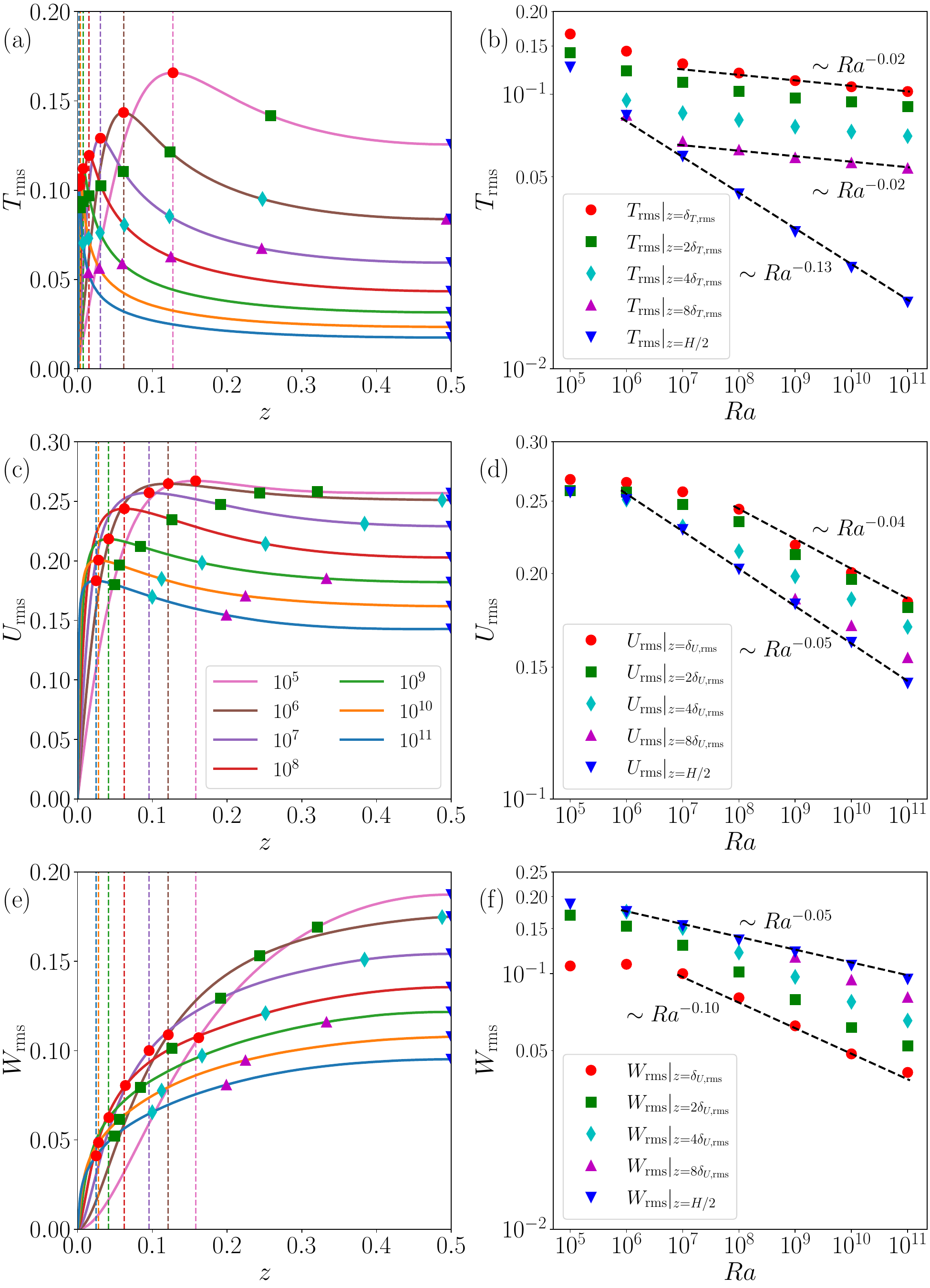}}
  \caption{(a) Profiles of temperature fluctuations, $T_{\mathrm{rms}}$, averaged across area $A$ and runtime $t$, for all Rayleigh numbers. The symbols indicate the values of these profiles at $z=\delta_{T,\mathrm{rms}}$ (red circles), $z=2\delta_{T,\mathrm{rms}}$ (green squares), $z=4\delta_{T,\mathrm{rms}}$ (cyan diamonds), $z=8\delta_{T,\mathrm{rms}}$ (purple triangles), and $z=H/2$ (blue inverted triangles). (b) Scaling of the indicated values of $T_{\mathrm{rms}}$ in (a) versus $Ra$. (c) Corresponding profiles of velocity fluctuations, $U_{\mathrm{rms}}$, with symbols marked in the same manner as in (a). The heights are now multiples of the momentum boundary region thickness, $\delta_{U,\mathrm{rms}}$. (d) Scaling of the indicated values of $U_{\mathrm{rms}}$ in (c) versus $Ra$. Similarly, the fluctuation profiles and scaling of the vertical velocity component, $u_z$, are shown in panels (e) and (f) respectively.}
\label{fig:rms_scale}
\end{figure}

\begin{table}
  \begin{center}
    \begin{tabular}{|l|c|c|c|c|c|c|c|}
    \hline
    \multirow{2}{*}{$Ra$} &
    \multirow{2}{*}{$\delta_{T,\mathrm{rms}}$} &
    \multirow{2}{*}{$\delta_{U,\mathrm{rms}}$} &
    \multicolumn{5}{|c|}{$T_{\mathrm{rms}}$, $\delta = \delta_{T,\mathrm{rms}}$} \\
    \cline{4-8}
    & & & $\delta$ & $2\delta$ & $4\delta$ & $8\delta$ & $H/2$ \\
    \hline
    $10^{ 5}$ & $1.28 \times 10^{-1}$ & $1.58 \times 10^{-1}$ & $0.166$ & $0.142$ &   -     &   -     & $0.126$ \\
    $10^{ 6}$ & $6.16 \times 10^{-2}$ & $1.21 \times 10^{-1}$ & $0.143$ & $0.121$ & $0.095$ & $0.084$ & $0.084$ \\
    $10^{ 7}$ & $3.06 \times 10^{-2}$ & $9.57 \times 10^{-2}$ & $0.129$ & $0.110$ & $0.085$ & $0.067$ & $0.059$ \\
    $10^{ 8}$ & $1.55 \times 10^{-2}$ & $6.26 \times 10^{-2}$ & $0.119$ & $0.102$ & $0.081$ & $0.063$ & $0.043$ \\
    $10^{ 9}$ & $7.47 \times 10^{-3}$ & $4.16 \times 10^{-2}$ & $0.112$ & $0.097$ & $0.076$ & $0.059$ & $0.032$ \\
    $10^{10}$ & $3.62 \times 10^{-3}$ & $2.80 \times 10^{-2}$ & $0.106$ & $0.094$ & $0.073$ & $0.056$ & $0.023$ \\
    $10^{11}$ & $1.80 \times 10^{-3}$ & $2.48 \times 10^{-2}$ & $0.102$ & $0.090$ & $0.070$ & $0.054$ & $0.017$ \\
    \hline
    \end{tabular}
    \caption{The boundary layer thicknesses $\delta_{T,\mathrm{rms}}$ and $\delta_{U,\mathrm{rms}}$ used to plot the scalings of figure~\ref{fig:rms_scale}, along with the values of $T_\mathrm{rms}$ at different heights. For the lowest $Ra$, the heights are sufficiently large such that $4\delta_{T,\mathrm{rms}}$ and above lie beyond $z=H/2$.} 
    \label{tab:trms}
  \end{center}
\end{table}

\begin{table}
  \begin{center}
    \begin{tabular}{|l|c|c|c|c|c|c|c|c|c|c|}
    \hline
    \multirow{2}{*}{$Ra$} &
    \multicolumn{5}{|c|}{$U_{\mathrm{rms}}$, $\delta = \delta_{U,\mathrm{rms}}$} &
    \multicolumn{5}{|c|}{$W_{\mathrm{rms}}$, $\delta = \delta_{U,\mathrm{rms}}$} \\
    \cline{2-11}
    & $\delta$ & $2\delta$ & $4\delta$ & $8\delta$ & $H/2$ & $\delta$ & $2\delta$ & $4\delta$ & $8\delta$ & $H/2$ \\
    \hline
    $10^{ 5}$ & $0.267$ & $0.258$ &   -     &   -     & $0.257$ & $0.107$ & $0.169$ &   -     &   -     & $0.187$ \\
    $10^{ 6}$ & $0.265$ & $0.257$ & $0.251$ &   -     & $0.251$ & $0.109$ & $0.153$ & $0.175$ &   -     & $0.175$ \\
    $10^{ 7}$ & $0.257$ & $0.247$ & $0.231$ &   -     & $0.229$ & $0.100$ & $0.129$ & $0.151$ &   -     & $0.154$ \\
    $10^{ 8}$ & $0.244$ & $0.235$ & $0.214$ &   -     & $0.203$ & $0.080$ & $0.101$ & $0.121$ &   -     & $0.136$ \\
    $10^{ 9}$ & $0.218$ & $0.212$ & $0.198$ & $0.185$ & $0.182$ & $0.063$ & $0.079$ & $0.097$ & $0.116$ & $0.122$ \\
    $10^{10}$ & $0.201$ & $0.196$ & $0.185$ & $0.170$ & $0.162$ & $0.049$ & $0.062$ & $0.078$ & $0.095$ & $0.108$ \\
    $10^{11}$ & $0.183$ & $0.180$ & $0.170$ & $0.154$ & $0.143$ & $0.041$ & $0.052$ & $0.066$ & $0.081$ & $0.095$ \\
    \hline
    \end{tabular}
    \caption{The values of $U_\mathrm{rms}$ and $W_\mathrm{rms}$ at different heights for all $Ra$, plotted in figure~\ref{fig:rms_scale}. The heights are multiples of $\delta_{U,\mathrm{rms}}$, which is listed in table~\ref{tab:trms}. Since $\delta_{U,\mathrm{rms}}$ tends to be much larger than $\delta_{T,\mathrm{rms}}$, we see that $8\delta_{U,\mathrm{rms}} > H/2$ even for $Ra$ as high as $10^8$.}
    \label{tab:urms}
  \end{center}
\end{table}

The planar-averaged profiles of the root mean square (rms) temperature and velocity fluctuations are shown in figure~\ref{fig:rms_scale}. The rms temperature fluctuations are calculated as $T_{\mathrm{rms}}(z) = \sqrt{\langle \theta^2({\bm x}, t) \rangle_{A,t}}$, where $\theta({\bm x}, t) = T({\bm x}, t) - \langle T(z) \rangle_{A,t}$. The maxima of the $T_{\mathrm{rms}}$ profiles are located close to the top and bottom plates, and the average distance between the maxima and their respective walls is a measure of the thermal boundary layer thickness, $\delta_{T,\mathrm{rms}}$. These heights are indicated with dashed lines along with the $T_{\mathrm{rms}}$ profiles in figure~\ref{fig:rms_scale}(a), colored according to their respective $Ra$ values. Similarly, the rms velocity fluctuations are calculated as $U_{\mathrm{rms}}(z) = \sqrt{\langle {\bm u}^2({\bm x}, t) \rangle_{A,t}}$, and the $U_{\mathrm{rms}}$ profiles also have their maxima close to the two plates, yielding a corresponding momentum (or viscous) boundary  thickness of $\delta_{U,\mathrm{rms}}$, see figure~\ref{fig:rms_scale}(c). Additionally, fluctuations in the vertical component of velocity are calculated as $W_{\mathrm{rms}}(z) = \sqrt{\langle u_z^2({\bm x}, t) \rangle_{A,t}}$, for which the maxima are always at the midplane of the convection cell at $z=H/2$. The $W_{\mathrm{rms}}$ profiles are plotted in figure~\ref{fig:rms_scale}(e) along with $\delta_{U,\mathrm{rms}}$ lines, similar to panel (c) of the same figure.

For a height-dependent analysis of the fluctuation profiles, we track the rms values at 5 heights, $z=\delta_{F,\mathrm{rms}}$, $2\delta_{F,\mathrm{rms}}$, $4\delta_{F,\mathrm{rms}}$, $8\delta_{F,\mathrm{rms}}$, and $H/2$, where $F=\{T,U\}$ for the temperature and velocity profiles respectively. These five locations are marked by filled symbols -- red circles, green squares, cyan diamonds, purple triangles, and blue inverted triangles, respectively, in figure~\ref{fig:rms_scale}. The values of fluctuation profiles corresponding to these heights are tabulated in tables~\ref{tab:trms} and~\ref{tab:urms} for temperature and velocity, respectively. Table~\ref{tab:trms} also lists the values of $\delta_{T,\mathrm{rms}}$ and $\delta_{U,\mathrm{rms}}$. The blank values in these tables correspond to those heights, which exceed the half height of $H/2$.

The scaling of the rms values at the chosen heights for all $Ra$ are plotted in figure~\ref{fig:rms_scale} in panels (b), (d) and (f) for temperature, total velocity and vertical velocity, respectively. At sufficiently high Rayleigh numbers of $Ra\geq 10^8$, all data follow a $Ra$-scaling law at all chosen heights and for all quantities. For the midplane, the scaling range is even extended to lower Rayleigh numbers. All near-wall data follow similar scalings for $T_{\rm rms}$ and $U_{\rm rms}$ with a very weak $Ra$-dependence. In case of $W_{\rm rms}$, the small scaling exponent increases from -0.10 at the edge of the boundary layer to -0.05 at the midplane. This scaling, together with the decreased amplitude towards the bulk, seems to be connected with the ongoing detachment of the thermal plumes and their subsequent rise into the bulk. The total velocity fluctuations (which include the horizontal velocity components) at the midplane display practically the same scaling as those above the boundary layer, again with a systematically decreased magnitude towards the bulk. Only the temperature fluctuations at the midplane have a smaller scaling exponent with -0.13 and thus a stronger $Ra$-dependence. See also \cite{Samuel2024} for a comparison with other previous investigations. This faster decay with $Ra$ indicates, that the bulk becomes increasingly well mixed for growing Rayleigh number. Plumes and plume clusters, that dominate the near-wall region, are more widely dispersed, which reduces the temperature fluctuations in the bulk more strongly. At the lowest Rayleigh numbers $Ra \lesssim 10^7$ the thermal plumes reach further into the bulk since the turbulent mixing is less developed. This seems to cause deviations from the scaling for most of the data.

\begin{figure}
  \centerline{\includegraphics[width=1.0\textwidth]{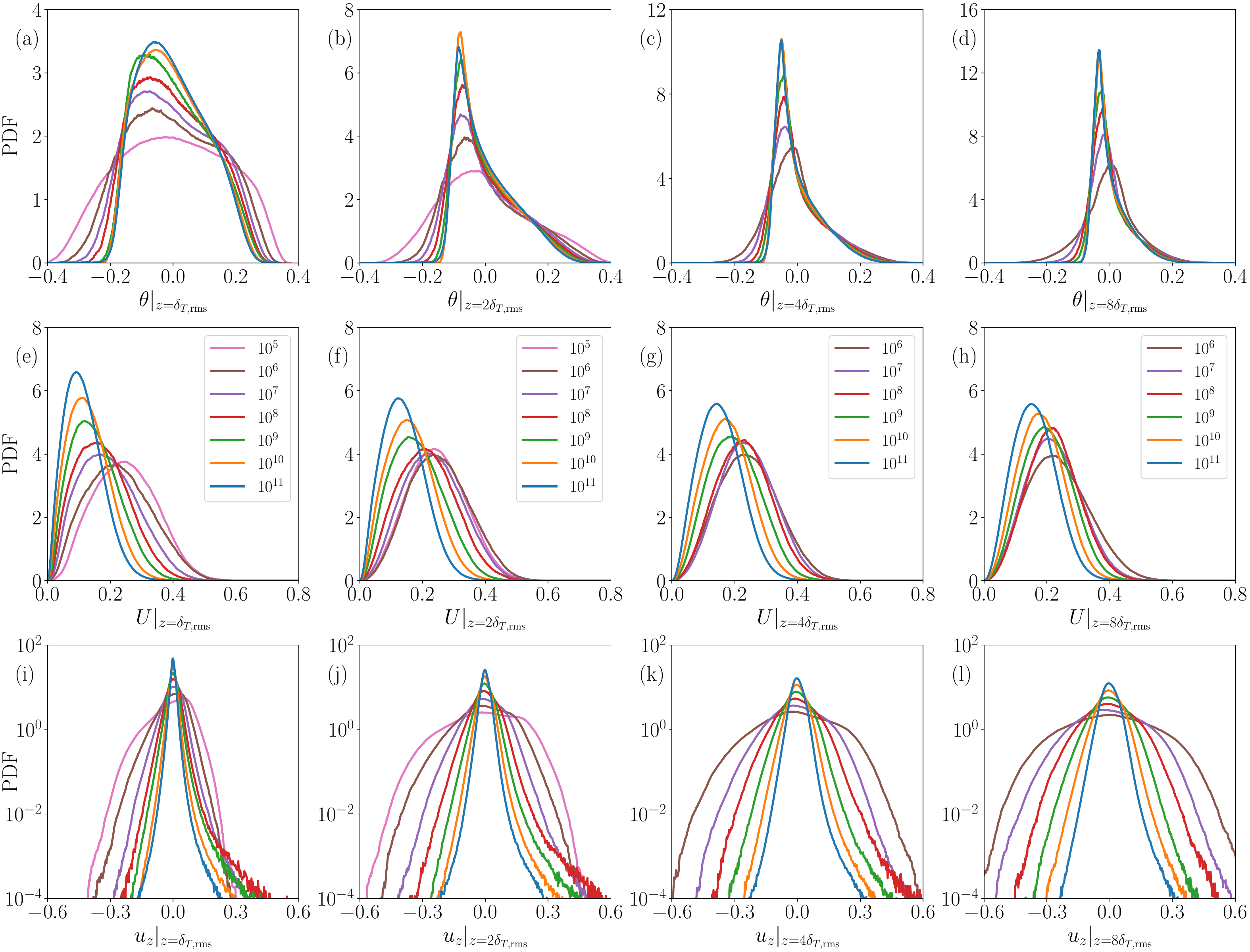}}
  \caption{Probability density functions (PDF) of the temperature fluctuations $\theta$ (top row), velocity magnitude $U=|{\bm u}|$ (middle row) and vertical velocity component $u_z$ (bottom row), sampled at horizontal cross-sectional $x$-$y$ planes at increasing heights of $z = \delta_T,~2\delta_T,~4\delta_T, \mathrm{and}~8\delta_T$ (left to right) for all $Ra$ values. The $y$ axis is logarithmic in the bottom row; it is linear in the upper and middle rows. The skewness of these distributions is tabulated in table~\ref{tab:skew}.}
\label{fig:vt_pdfs}
\end{figure}

\begin{table}
  \begin{center}
    \begin{tabular}{|l|r|r|r|r|r|r|r|r|}
    \hline
    \multirow{2}{*}{Ra}  & \multicolumn{4}{|c|}{$\theta$} & \multicolumn{4}{|c|}{$W$} \\
    \cline{2-9}
              & \cc{$\delta_T$} & \cc{$2\delta_T$} & \cc{$4\delta_T$} & \cc{$8\delta_T$} & \cc{$\delta_T$} & \cc{$2\delta_T$} & \cc{$4\delta_T$} & \cc{$8\delta_T$} \\
    \hline
    $10^{ 5}$ & -0.023 &  0.359 & \cc{-} & \cc{-} & -0.620 & -0.228 & \cc{-} & \cc{-} \\
    $10^{ 6}$ &  0.074 &  0.607 &  0.640 &  0.030 & -0.564 & -0.208 & -0.046 &  0.015 \\
    $10^{ 7}$ &  0.149 &  0.685 &  0.916 &  0.807 & -0.378 & -0.108 &  0.011 &  0.028 \\
    $10^{ 8}$ &  0.209 &  0.792 &  1.104 &  1.176 & -0.022 &  0.019 &  0.040 &  0.017 \\
    $10^{ 9}$ &  0.333 &  0.863 &  1.183 &  1.309 &  0.169 &  0.099 &  0.053 &  0.001 \\
    $10^{10}$ &  0.295 &  0.893 &  1.220 &  1.370 &  0.324 &  0.193 &  0.116 &  0.026 \\
    $10^{11}$ &  0.289 &  0.811 &  1.141 &  1.316 &  0.242 &  0.188 &  0.170 &  0.086 \\
    \hline
    \end{tabular}
    \caption{Skewness of probability density functions of the fluctuating fields, $\theta$ and $W$, at different heights for all $Ra$. The distributions are plotted in figure~\ref{fig:vt_pdfs}.}
    \label{tab:skew}
  \end{center}
\end{table}


\subsection{Probability density function of velocity and temperature fluctuations}
\label{sec:pdfs}

The probability density functions (PDFs) of velocity and temperature fluctuations show marked variations with Rayleigh number -- the distributions tend to become narrower with increasing $Ra$, as seen in figure~\ref{fig:vt_pdfs}. In the figure, the PDFs of the temperature fluctuation $\theta$ are shown for cross-sectional planes taken at 4 different heights from the bottom plate, namely $z=\delta_{T,\mathrm{rms}}$, $2\delta_{T,\mathrm{rms}}$, $4\delta_{T,\mathrm{rms}}$, and $8\delta_{T,\mathrm{rms}}$, in the upper row. With increasing distance from the wall the temperature field gets increasingly mixed which manifests in the narrower distributions.

The corresponding PDF of the velocity magnitude $U=|{\bm u}|$ and vertical velocity component $u_z$ are presented in the middle and bottom rows respectively. Note that unlike in the previous section, where the velocity fluctuations were sampled at multiples of $\delta_{U,\mathrm{rms}}$, the PDFs are calculated here also at planes corresponding to $\delta_{T,\mathrm{rms}}\approx \delta_T$ for comparison between both fields. The thermal boundary layer thickness is our reference scale for comparisons of the statistics of both fields. In \cite{Samuel2024}, we showed that the ratio of $\delta_{U,{\rm rms}}/\delta_{T,{\rm rms}}$ grows with $Ra$, see table \ref{tab:trms}. The PDFs for temperature are distinctly different at the edge of the thermal boundary layer, as compared with the other heights. The range of bins of the distributions is kept constant for all the heights to clearly indicate their narrowing or widening, as we transition from the wall to the bulk. 

We also see that the support of the PDF of the vertical velocity component $u_z$, which again is not rescaled by the rms value, decreases with growing $Ra$. The reason for this observation is connected to the fact that we probe at heights taken in units of $\delta_T$ and that the ratio of the fluctuation scale thicknesses of $\delta_{U,{\rm rms}}$ to $\delta_{T,{\rm rms}}$ increases with $Ra$.

Table \ref{tab:skew} summarizes the skewness values of the distributions from figure \ref{fig:vt_pdfs}. The following trends can be observed. For $Ra\ge 10^8$, the skewness of the PDF of the temperature fluctuations with respect to $Ra$ remains in a similar range; it, however, increases with growing wall distance at fixed $Ra$. This is a manifestation of the plume clustering, which we analyzed in Shevkar et al. (2025). Contrarily, the skewness of the PDF of the vertical fluctuations decreases at fixed $Ra$ with increasing wall distance. We interpret this as a manifestation of the enhanced vigor of turbulence away from the wall with growing Rayleigh number. This symmetrizes the vertical velocity distribution.
\begin{figure}
  \centerline{\includegraphics[width=1.0\textwidth]{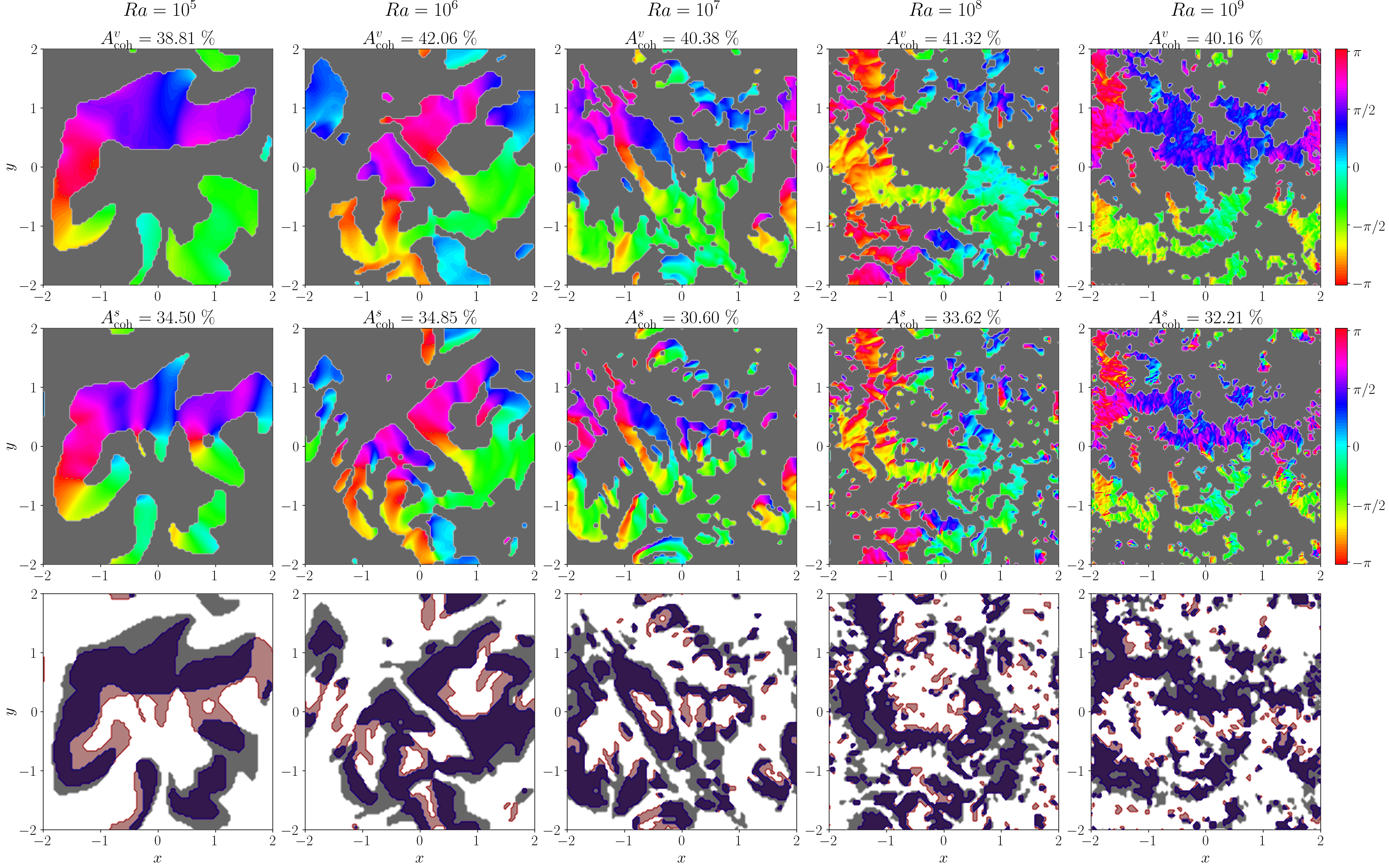}}
  \caption{Decomposition of the boundary region flow field into shear-dominated coherent sections and plume-dominated incoherent sections at the thermal boundary layer height $z = \delta_T$. Two methods are compared for $Ra$ ranging from $10^5$ to $10^9$ - thresholding the rms horizontal velocity, $U_h$ (top row), and the horizontal shear-stress, $S_h$ (middle row). The corresponding area fractions of the coherent regions are also indicated as $A^v_\mathrm{coh}$ and $A^s_\mathrm{coh}$ respectively. The cyclic colormap within the coherent regions indicates the angle of the flow in the wall-parallel plane, ranging from $-\pi$ to $\pi$. The agreement between the two methods is highlighted in the bottom row, where the coherent regions are plotted for velocity threshold in grey, shear-stress threshold in red, and the overlapping regions of both methods, $A^v_\mathrm{coh} \cap A^s_\mathrm{coh}$, in dark blue. The ratio between the overlapping area and $A^s_\mathrm{coh}$ is are 0.72, 0.78, 0.80, 0.80 and 0.84 for $Ra = 10^5, 10^6, 10^7, 10^8$ and $10^9$ respectively.}
\label{fig:decomp_comp}
\end{figure}

\section{Nusselt number in coherent and incoherent boundary layer regions}
\label{sec:decompose}

The flow in the boundary region of a plane-layer unconfined RBC is understood to be a patchwork of shear-dominated (coherent) and plume-dominated (incoherent) regions. The in-plane horizontal velocity, $\mathbf{u_h} = (u_x, u_y)$, was shown to be a reliable measure for identifying such coherent and incoherent areas in \cite{Samuel2024}. This is done by breaking up the 2D flow-field in the $x$-$y$ plane at the edge of the thermal boundary layer, $z=\delta_T$, into $100 \times 100$ disjoint tiles of area $A_i = A/N_{\rm tiles}$. Here, $N_{\rm tiles}$ is the number of ``square tiles'' that cover the whole cross-section $A$. The area-averaged planar rms velocities are computed for individual tiles, as well as across the whole plane. In the following, we take
\begin{equation}
U^i_h = \sqrt{\langle u_x^2 + u_y^2 \rangle_{A_i}} \quad \mathrm{and} \quad U_h = \sqrt{\langle u_x^2 + u_y^2 \rangle_A}.
\label{eq:urms_tiles}
\end{equation}
The tiles with $U^i_h \geq U_h$ indicate shear-dominated regions, whereas the tiles with $U^i_h < U_h$ are labeled as plume-dominated. Alternatively, one can also take the shear stress field at the no-slip plates to isolate shear dominated sections from the rest. We investigate this alternative criterion by focusing on the horizontal shear-stress field, $\mathbf{s_h} = (s_x, s_y)$, where
\begin{equation}
s_x = \left|\frac{\partial u_x}{\partial z}\right| \quad \mathrm{and} \quad s_y = \left|\frac{\partial u_y}{\partial z}\right|.
\label{eq:s_xy}
\end{equation}
Similar to the velocity-based criterion, the local and global area-averaged rms shear stress, namely $S^i_h$ and $S_h$ respectively, are calculated over disjoint tiles.

\begin{figure}
  \centerline{\includegraphics[width=1.0\textwidth]{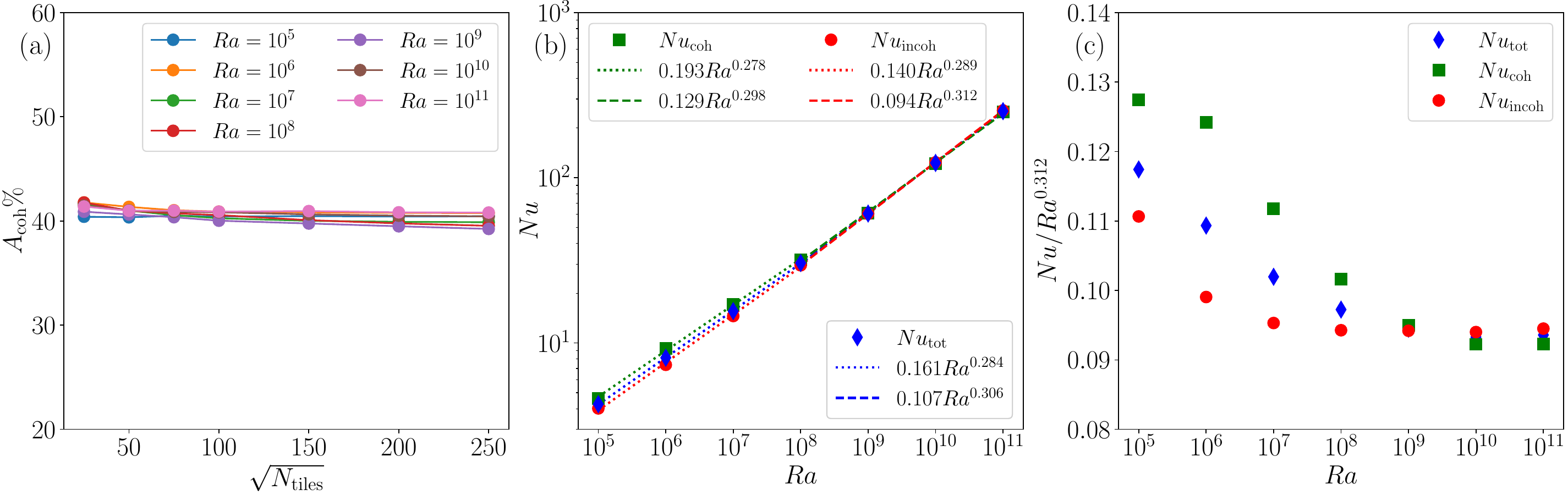}}
  \caption{Area fraction and Nusselt number scaling of coherent regions of the boundary region for $10^5 \leq Ra \leq 10^{11}$. (a) Area fraction $A_{\rm coh}$ versus $Ra$ for different decompositions of the whole plane, here given by the number $\sqrt{N_{\rm tiles}}$. (b) Nusselt number scaling $Nu(Ra)$ in double logarithmic plot conditioned to coherent and incoherent area fractions as well as for total area. The dotted lines are fits for $10^5\le Ra\le 10^8$, the dashed ones for $10^8\le Ra\le 10^{11}$ (c) Compensated linear-logarithmic replot of the data of panel (b). The contribution to global heat transport from the shear-dominated regions decreases with increasing $Ra$. In the high-Rayleigh-number regime for $Ra>10^8$, scaling of $Nu(Ra)$ is closest to $1/3$ in the plume dominated regions, whereas the shear-dominated regions display a smaller scaling exponent.}
\label{fig:coh_incoh_nu}
\end{figure}

Figure~\ref{fig:decomp_comp} compares these two methods of decomposing the horizontal velocity field at the edge of the thermal boundary layer, $z = \delta_T$, for four Rayleigh numbers ranging from $10^5 \leq Ra \leq 10^9$. The greyed-out patches represent the regions where $U^i_h < U_h$ (incoherent flow). The underlying cyclic colormap denotes the angle of the direction of the flow with respect to the positive $x$-axis. Areas with a smoothly varying color-gradient therefore represents a local patch with coherent unidirectional flow, whereas areas with rapidly fluctuating colors represent chaotic regions with plume-upwelling (covered by the grey patches denoting $A_\mathrm{incoh}$). Both methods give a consistent and similar distribution of coherent and incoherent regions. The area fraction of the region with coherent flow is the ratio between the summed area of shear-dominated tiles and the total cross-sectional area, $A$. The area fractions calculated from $\mathbf{u_h}$ and $\mathbf{s_h}$ are denoted $A^v_\mathrm{coh}$ and $A^s_\mathrm{coh}$ respectively. The instantaneous values of the area fractions are indicated for the snapshots in figure~\ref{fig:decomp_comp}, and these are similar to the average coherent area fraction of $A_{\rm coh}\approx 40$\% observed earlier in \cite{Samuel2024}, see figure \ref{fig:coh_incoh_nu}(a). Furthermore, the overlap area of the coherent flow regions computed from the two methods, $A^v_\mathrm{coh} \cap A^s_\mathrm{coh}$, is indicated by the dark blue regions in the bottom row of the figure. The degree of agreement between the two methods is calculated as $(A^v_\mathrm{coh} \cap A^s_\mathrm{coh})/A^s_\mathrm{coh}$. This agreement improves with $Ra$, improving from 0.72 at $Ra=10^5$ to 0.84 at $Ra=10^9$. The values at intermediate $Ra$ are listed in the caption.

The division into coherent shear-dominated and incoherent plume-dominated regions can be also established on the basis of the temperature field, which we did in \cite{Shevkar2025}. Regions with $T>T_{\rm rms}(z)$ can be assigned as plume-dominated regions; the complementary area fraction would be shear-dominated or impact-dominated. The area fraction $A_{\rm incoh}$, which is obtained in this way, is 45\% at $z=\delta_T$, again relatively independent of the Rayleigh number for the whole range from $Ra=10^5$ to $10^{11}$. This would correspond to $A_{\rm coh}\approx 55\%$, which is of the same order of magnitude as in figure \ref{fig:coh_incoh_nu}(a). Important is to our view not the specific area fraction value, but the independence of $Ra$ which is obtained in different ways for different fields.

As shown in \cite{Shevkar2025} and discussed in the introduction, the thermal plume network gets ever finer with increasing Rayleigh number. A mean plume spacing scale decreases with approximately $Ra^{-1/3}$ and thus stays to a good approximation in a fixed ratio with the mean thermal boundary layer thickness $\delta_T$. The latter scale sets the stem thickness of the detaching thermal plumes. This constant ratio seems to be the reason for the occupation of the same area fraction with growing $Ra$. The network gets ever finer, but the area fraction that it occupies, remains the same.

We also stress on the independence of the observed area-fractions on the tile size used for decomposition. In figure~\ref{fig:coh_incoh_nu}(a), we compare the $A_\mathrm{coh}$ percentage values for different numbers of tiles for different $Ra$. The higher $Ra$ cases are more affected by coarse-grained tiling than the lower $Ra$ data. Nevertheless, the area-fractions remain robust and do not vary too much from the 40$\%$ value.

Finally, we investigate the effect of shear flow on local heat transfer by computing the corresponding temperature gradients at the wall. Consequently, we define the two locally area-averaged Nusselt numbers for these two regions as
\begin{equation}
Nu_{\mathrm{coh}}=-\left \langle \frac{\partial T}{\partial z} \bigg|_{z=0}\right \rangle_{A \in A^v_{\mathrm{coh}},t} \qquad
Nu_{\mathrm{incoh}}=-\left \langle \frac{\partial T}{\partial z} \bigg|_{z=0}\right \rangle_{A \in A^v_{\mathrm{incoh}},t}.
\end{equation}
\begin{table}
  \begin{center}
    \begin{tabular}{|l|c|c|c|}
    \hline
    $Ra$ & $Nu_{\mathrm{tot}}$ & $Nu_{\mathrm{coh}}$ & $Nu_{\mathrm{incoh}}$ \\
    \hline
    $10^{ 5}$ & $  4.26 \pm 0.18$ & $  4.63 \pm 0.24$ & $  4.02 \pm 0.21$ \\
    $10^{ 6}$ & $  8.14 \pm 0.18$ & $  9.25 \pm 0.32$ & $  7.38 \pm 0.21$ \\
    $10^{ 7}$ & $ 15.58 \pm 0.26$ & $ 17.07 \pm 0.48$ & $ 14.56 \pm 0.22$ \\
    $10^{ 8}$ & $ 30.47 \pm 0.44$ & $ 31.83 \pm 0.82$ & $ 29.54 \pm 0.38$ \\
    $10^{ 9}$ & $ 60.73 \pm 0.65$ & $ 61.04 \pm 0.97$ & $ 60.54 \pm 0.80$ \\
    $10^{10}$ & $122.96 \pm 1.24$ & $121.60 \pm 1.55$ & $123.92 \pm 1.41$ \\
    $10^{11}$ & $253.04 \pm 1.52$ & $249.50 \pm 2.29$ & $255.53 \pm 2.16$ \\
    \hline
    \end{tabular}
    \caption{The values of Nusselt number $Nu$ in the plume-dominated ($Nu_{\mathrm{incoh}}$) and shear-dominated regions ($Nu_{\mathrm{coh}}$) compared with the one taken over the whole horizontal cross-section area ($Nu_{\mathrm{tot}}$). The data are plotted in figure~\ref{fig:coh_incoh_nu}.}
    \label{tab:cohincoh}
  \end{center}
\end{table}
The scaling of $Nu_{\mathrm{coh}}$ and $Nu_{\mathrm{incoh}}$ are shown in figure~\ref{fig:coh_incoh_nu}(b) along with their corresponding fits, both for the moderate ($Ra \le 10^8$) and higher ($Ra\ge 10^8$) Rayleigh numbers. These values are also summarized in table~\ref{tab:cohincoh}. Interestingly, the shear-dominated sections have a higher heat-transfer at low to moderate $Ra$ values up to $Ra = 10^9$. This trend reverses at $Ra\geq 10^{10}$, where the plume-dominated regions contribute to a greater share of the overall heat transport. This is highlighted by the compensated plot in figure~\ref{fig:coh_incoh_nu}(c), where $Nu$ has been compensated by the highest exponent obtained from the power-law fits, $Ra^{0.312}$. We note that similar decompositions of local heat transfer near the wall were performed in the case of two-dimensional RBC by Zhu et al. \cite{Zhu:PRL2018} and He et al. \cite{He2024} for periodic side boundaries and solid side walls, respectively. Plume-dominated regions were further broken down into plume ejection and plume impact regions and differences in the local turbulent heat transfer and the mean profiles of velocity and temperature were identified.

\section{Long-term evolution of convection under different imposed shear flows}
\label{sec:shear}

We now consider the effect of different forms of shear flow imposed on top of natural plane-layer convection. Our goal here is to delve deeper into the effect of plume- and shear-dominated regions on the overall turbulent heat transfer in RBC. Furthermore, we explore the existence or lack thereof of the well-known logarithmic velocity profile, as expected for a turbulent boundary layer under strong shearing winds from the bulk. For all experiments described in this section, the shear flow is imposed along an arbitrarily chosen direction, which is without loss of generality the $x$-direction. We retain the $\Gamma = 4$ domain with periodic horizontal extents, and fix the Rayleigh number and Prandtl number as $Ra=10^9$ and $Pr=0.7$, respectively.

The strength of the imposed flow is gradually increased over the next sub-sections. A decaying sinusoidal mode shear flow at different amplitudes is first considered in \sref{sec:initial}. The focus of this analysis is to test whether different turbulent macrostates coexist, as motivated in the introduction. A steady sinusoidal forcing is then applied to amplify the large-scale circulations, to generate a uni-directional mean flow, and to potentially trigger a logarithmic velocity profile in \sref{sec:forced}. Finally, a steady pressure-gradient is applied to generate a shear-dominated flow in \sref{sec:pgrad}, which is compared to the one in section 5.2. The Rayleigh number is fixed to $Ra=10^9$ throughout this section. On the one hand, it is sufficiently high enough to develop turbulence that is sensitive to different initial conditions; on the other hand, it allows us to obtain the long-term integration results with reasonable numerical effort.

\begin{figure}
  \centerline{\includegraphics[width=1.0\textwidth]{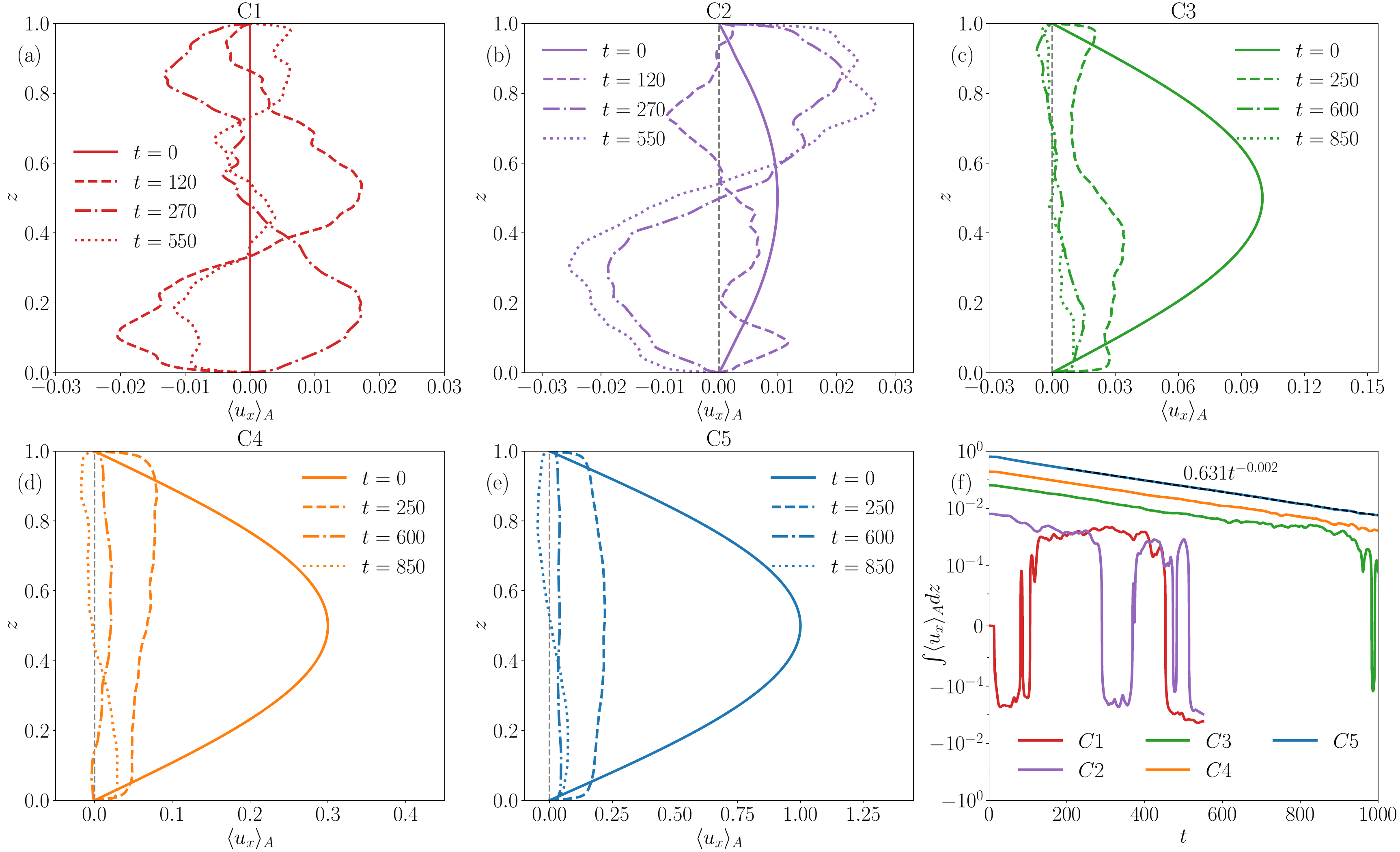}}
  \caption{Snapshots of the planar-averaged mean vertical profiles of the $x$-component of velocity at 4 instances of non-dimensional time, and the rates of their decay. Cases C1 through C5 are shown in panels (a) to (e), respectively. The last panel shows the temporal variation of the area under the velocity profile integrated along the vertical $z$-direction. The area is positive when there is a net mean flow in $x$-direction and vice-versa. See also table \ref{tab:shear} for the parameters of the runs.}
\label{fig:profile_decay}
\end{figure}

\subsection{Decay of an initially applied sinusoidal shear flow mode}
\label{sec:initial}

For this study, a finite-amplitude unidirectional shear flow is initially imposed along the $x$-direction. Over a sufficiently long duration, this initial flow will decay and the buoyancy-driven convective flow will dominate. The initial velocity condition therefore has a profile of $u_x(z) = \tilde{A}_i \sin(\pi z)$. This flow is added to the standard initial condition, the infinitesimally perturbed linear temperature profile.

To clearly quantify a possible susceptibility of a long-term evolution to different macrostates of the RBC flow, an additional initial {\em finite-amplitude} flow mode is applied. We consider 5 cases with different values of initial flow amplitude $\tilde{A}_i$, named C1, C2, C3, C4 and C5, with $\tilde{A}_i = 0$, 0.01, 0.1, 0.3 and 1.0 respectively. Of these, C1 represents the standard RBC case. All five cases have a linear temperature profile $T(z) = 1 - z$ as the initial condition for the temperature. The initial velocity profiles of the cases listed above can be observed by the solid lines in panels (a) to (e), respectively, of figure~\ref{fig:profile_decay}.

Figure~\ref{fig:profile_decay}(f) traces the temporal variation of the mean flow along $x$-direction, which is calculated as $\int_0^1 \langle u_x \rangle_A dz$. This is a flow rate. For the cases with weak or no initial shear flow, C1 and C2, we observe a randomly fluctuating mean flow indicating that the convective current dominates across the domain. Nevertheless, in case C2 it takes a short duration for the initial shear flow mode to decay. The buoyancy-driven flow takes over only after approximately 200 free-fall time units. The cases C3, C4 and C5 have a sufficiently stronger initial sinusoidal flow amplitude, such that it requires more than 550 free-fall time units for a full decay. Consequently, these cases were run for longer total integration times to quantify the time it takes for the signature of initial shear flow to disappear completely. The temporal decay of the shear flow mode follows a consistent power law of $t^{-0.002}$, with prefactors that increase with the amplitude $\tilde{A}_i$ of the initial sinusoidal profile. We observe that C3 begins to deviate from this power law after approximately 650 free-fall times. After nearly 950 units, the mean flow begins to fluctuate around zero mean, like in cases C1 and C2. Furthermore, cases C4 and C5 require far longer times than 1000 units for the flow to recover a completely buoyancy-dominated dynamics. In all cases, the shear mode, which was initially excited, is not sustained by the nonlinear couplings of the degrees of freedom of the turbulent flow. For a pure sine mode $u_x(z) = \tilde{A}_i \sin(\pi z)$ subjected to viscous effects alone, the velocity field decays as $u_x(z, t) = \tilde{A}_i \sin(\pi z) e^{-\tilde\nu \pi^2 t}$. This corresponds to a decaying flow rate of $(2\tilde{A}_i/\pi) e^{-\tilde\nu \pi^2 t}$. Here, $\tilde\nu=\sqrt{Pr/Ra}$, the dimensionless kinematic viscosity. This decay rate is significantly slower than the observed power law of $t^{-0.002}$, demonstrating that the convective turbulence contributes overwhelmingly to the overall decay of the flow rate.

\begin{figure}
  \centerline{\includegraphics[width=1.0\textwidth]{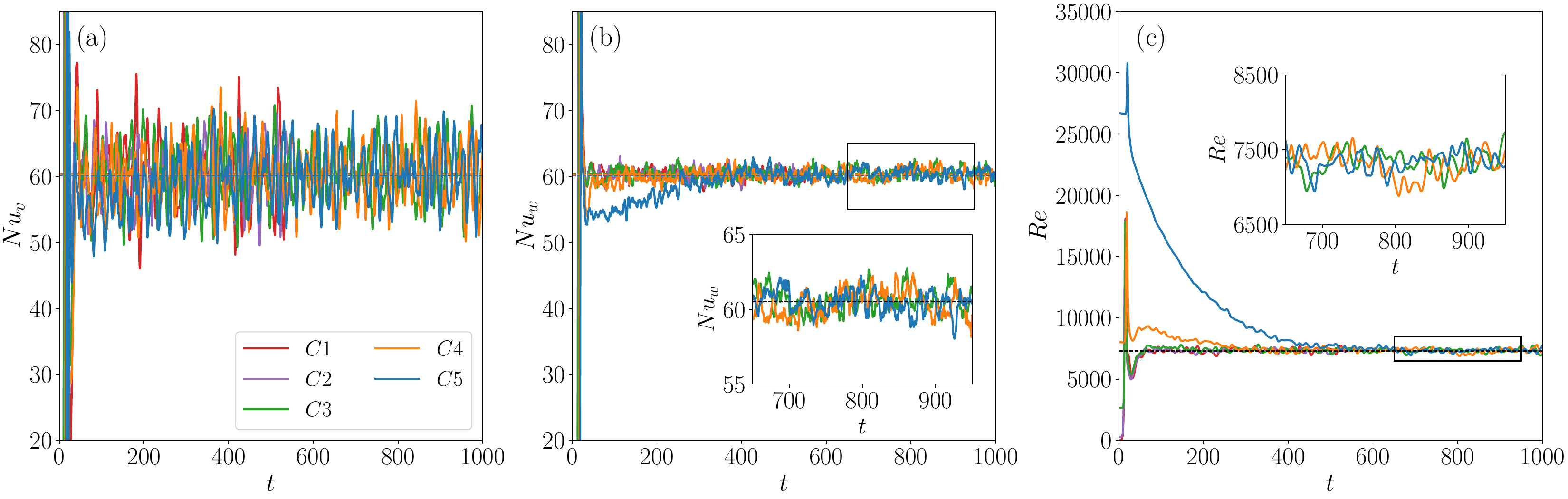}}
  \caption{Time-series of the (a) volume-averaged Nusselt number, $Nu_v$, (b) area-averaged wall Nusselt number, $Nu_w$, and (c) Reynolds number, $Re$, for the 5 cases with varying amplitudes of initial sinusoidal forcing at $Ra=10^9$. The large fluctuations in $Nu_v$ mask the lower heat transfer in the strongly shear-dominated case, which is made visible by the $Nu_w$ plot. As the effect of the strong shear flow subsides, the Nusselt number increases to the expected value of 60.5 for $Ra=10^9$~\cite{Samuel2024}.}
\label{fig:sinic_ts}
\end{figure}

\begin{figure}
  \centerline{\includegraphics[width=1.0\textwidth]{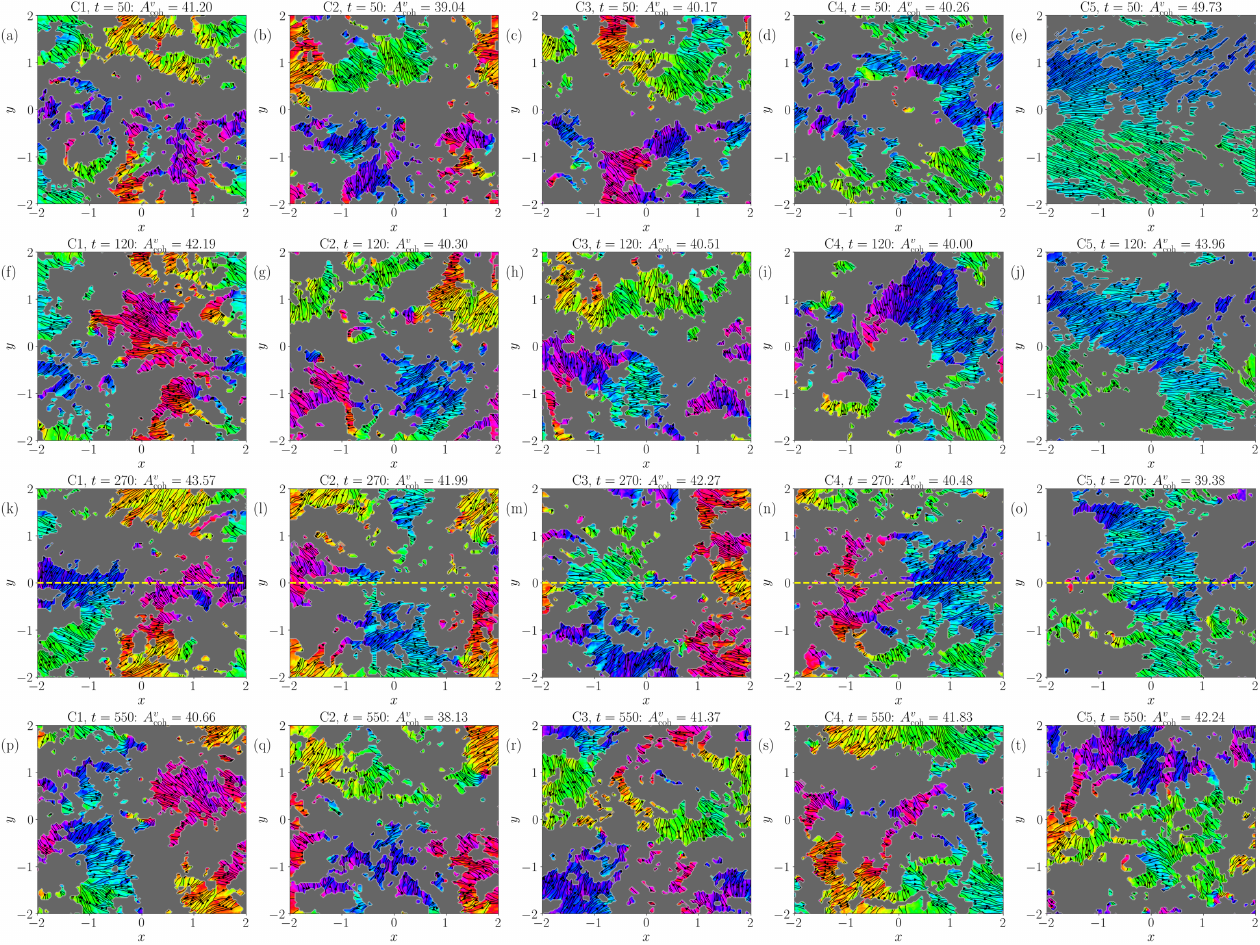}}
  \caption{Decomposition of flow field at the thermal boundary layer, $z=\delta_T$, into shear and plume-dominated sections with the same color mapping as in figure~\ref{fig:decomp_comp}. The five columns from left to right show cases C1, C2, C3, C4 and C5 respectively. The four rows from top to bottom are snapshots of each case at $t = 50$, 120, 270 and 550 respectively. Interestingly, there are significant plume dominated sections even in the cases with strong shear flow. This is explained by the overall $U_\mathrm{rms}$ threshold for splitting the flow being elevated when there is a background flow. The dashed yellow line in the third row indicates the vertical planes shown in figure~\ref{fig:uxField}.}
\label{fig:area_frac}
\end{figure}

\begin{figure}
  \centerline{\includegraphics[width=0.95\textwidth]{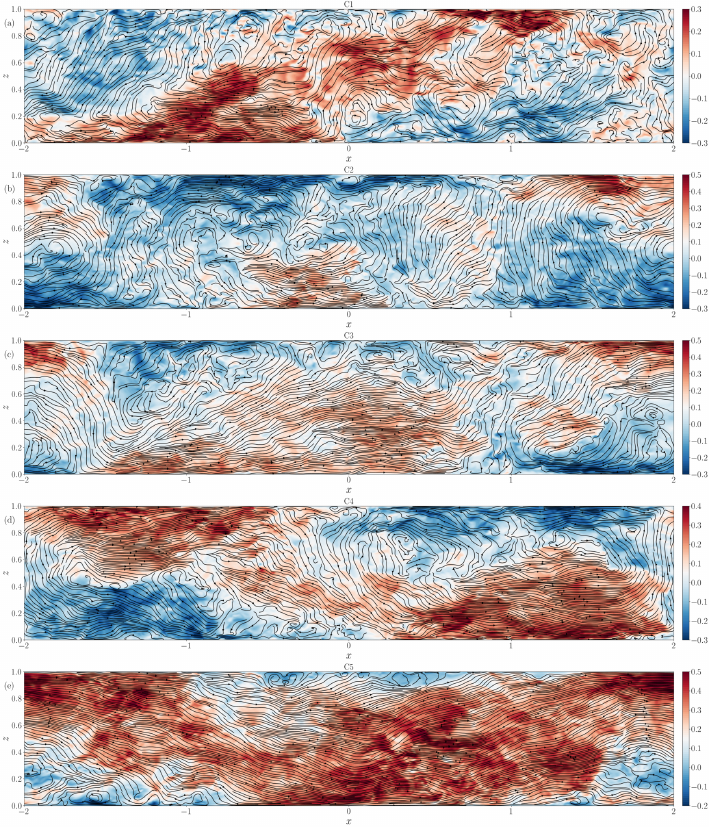}}
  \caption{Structure of flow field on the vertical $xz$-plane at $y=0$ for cases C1 through C5 at $t=270$. Although there is a strong initial shear flow in C5, the mean flow adopts a wavy-structure, alternatively grazing the top and bottom plates, allowing for pockets of reverse flow, driven by the thermal plumes. This explains the patches of reverse flow observed on the plate in figure~\ref{fig:area_frac} and the relatively low area fraction of shear-dominated patches.}
\label{fig:uxField}
\end{figure}

The globally averaged dimensionless quantities Reynolds number, $Re$, and Nusselt number, $Nu$, represent the total momentum and heat flux, respectively. They were defined in \eref{eq:NuRe}. The intensity of turbulence increases with $Ra$ and consequently so does the $Re$. We now consider two approaches for calculating the Nusselt number -- from the wall temperature gradient and from the convective transport of temperature -- which are represented as $Nu_w$ and $Nu_v$ respectively,
\begin{equation}
Nu_w=-\frac{\partial \langle T\rangle_{A,t}}{\partial z}\bigg|_{z=0} \quad\mbox{and}\quad
Nu_v=1+\sqrt{Ra Pr}\langle u_z T\rangle_{V,t}.
\end{equation}
Nusselt number $Nu_w$ corresponds to $Nu$ in \eref{eq:NuRe}. The notation has been changed to distinguish this number from $Nu_v$. The temporal variation of these dimensionless quantities is plotted in figure~\ref{fig:sinic_ts}. The volume-averaged heat flux typically tends to fluctuate greatly as observed in figure~\ref{fig:sinic_ts}(a). This masks the differences in heat transfer across the five cases. Nevertheless, $Nu_w$ in (b), which varies to a lesser degree, highlights the differences between the shear-dominated and plume-dominated states of the flow. In case C5, for the first 200 free-fall times when the system has a significant mean flow, the Nusselt number is noticeably lower. The Reynolds number $Re$ in panel (c) indicates higher values for stronger mean flows as expected, and all cases converge to the common value of approximately 7350 within the error bars.

Now, we apply again the decomposition of the flow in the near-wall region into coherent and incoherent regions, as already presented in \sref{sec:decompose}, to the shear-flow cases at times $t =$ 50, 120, 270 and 550, all of which lie within the range of durations common to all five cases. Figure~\ref{fig:area_frac} shows this decomposition for cases C1 to C5 arranged column-wise from left to right, and at increasing units of time from top to bottom. The underlying contours show the flow angle, similar to figure~\ref{fig:decomp_comp}, and the streamlines illustrate the velocity field. For C5 at $t=50$, the area fraction of the coherent flow region is 50\% and not 40\% as in all other cases. One would nominally expect a higher area fraction for such a strong initial mean flow at such an early stage of the evolution of flow. All cases still seem to have a significant plume-dominated region despite the imposed shear flow. One reason for this is the elevated threshold value of $U_\mathrm{rms}$ due to background flow.

Figure~\ref{fig:uxField} provides a more compelling explanation for the smaller area fraction even with applied shear. The figure shows the vertical cross-section ($x$-$y$ plane) of the flow at half of the spanwise extension, $y=0$, for cases C1 through C5 arranged from top to bottom. All snapshots are taken at time $t=270$. We see that the flow gets organized into a wavy structure that grazes the top and bottom plates in an alternating way as it moves from left to right. This reorganization of RBC flow under imposed shear has also been observed in the case with Couette-flow shearing~\cite{Blass:JFM2020,Blass:JFM2021}. As a result, the flow near the walls has alternating regions of shear flow with pockets of buoyancy-driven recirculating flow in between, bringing down the area fraction of coherent flow.

\begin{figure}
  \centerline{\includegraphics[width=0.95\textwidth]{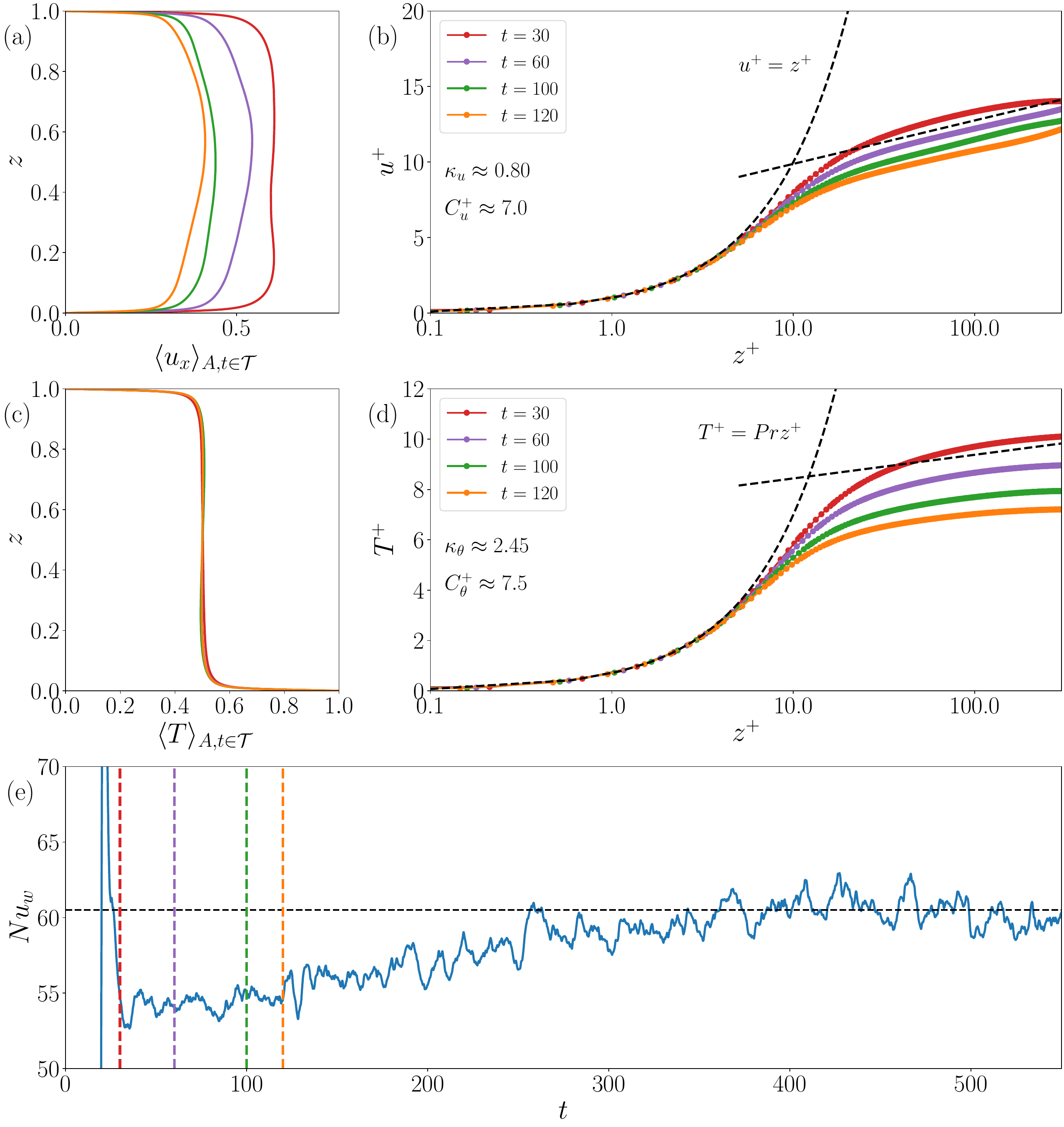}}
  \caption{(a) Planar-averaged profiles of $x$-component of velocity, $\langle u_x \rangle_{A,t \in \mathcal{T}}$, for case C5, at four time instants, $t =$ 30, 60, 100 and 120. Here, $\mathcal{T}$ is a short time-averaging window of 10 free-fall times centered around the respective $t$ values. (b) The corresponding rescaled velocity profiles in wall units along with the viscous-sublayer and log-law regimes indicated with dashed black lines. (c) Planar-averaged profiles of temperature, $\langle T \rangle_{A,t \in \mathcal{T}}$, again for case C5, at the same instants as in panels (a) and (b). (d) The corresponding rescaled temperature profiles in wall units. (e) The four instants highlighted in the previous panels are indicated with dashed vertical lines in the time-series of $Nu_w$ to demonstrate the impact on the global heat transport.}
\label{fig:loglaw}
\end{figure}

From the planar-averaged profiles of the streamwise velocity component in figure~\ref{fig:profile_decay} we observe that as the high-amplitude initial flow decays for case C5 in panel (e), the profile at $t=250$ appears to have a thin near-wall mean profile similar to that observed in wall-bounded shear flows. We therefore focus on the initial decay of C5 in figure~\ref{fig:loglaw}, where we investigate potential signatures of logarithmic mean velocity profile. Four snapshots are chosen at $t=$ 30, 60, 100 and 120; the corresponding data are plotted in red, purple, green and orange respectively. To smooth out rapid fluctuations in the profiles, a short averaging window of 10 free-fall times, $\mathcal{T}$, is used to compute time averages. From the velocity profiles $\langle u_x \rangle_{A,t \in \mathcal{T}}$ shown in (a), we observe the D-shaped profile at $t=30$.

Similar to the analysis of free-stream boundary layers \cite{Pope2000}, we compute the friction velocity, $u_{\tau}$, and the inner length scale, $l^+$, as
\begin{equation}
u_\tau=\sqrt{\frac{\tau_w}{\rho}}, \quad\quad
l^+=\frac{\nu}{u_\tau}, \quad\quad \mbox{where} \quad\quad
\tau_w=\mu \frac{d\langle u_x \rangle_{A,t \in \mathcal{T}}}{dz}\biggr|_{z=0}.
\label{eq:wall_scale}
\end{equation}
Here $\mu$ and $\rho$ are the dynamic viscosity and mass density, respectively, so that $\nu = \mu/\rho$. Using $u_{\tau}$ and $l^+$, the velocity and length scales are rescaled to obtain the profiles in terms of wall units, $u^+ = \langle u_x \rangle_{A,t \in \mathcal{T}}/u_\tau$, and $z^+ = z/l^+$. The resulting profiles are plotted in figure~\ref{fig:loglaw}(b) along with the viscous sub-layer curve, which is given by $u^+ = z^+$. The logarithmic law of the wall follows to 
\begin{equation}
u^+(z^+) = \frac{1}{\kappa_u} \log{z^+} + C^+_u\,.
\label{eq:wall_logu}
\end{equation}
The von K\'{a}rm\'{a}n constant $\kappa_u$ and the offset $C^+_u$ are expected to be approximately 0.4 and 5.5, respectively, for standard wall-bounded turbulent shear flows, such as pipe flow, plane Poiseuille flow, or plane Couette flow \cite{Pope2000}. Although the viscous-sublayer is well-resolved --- each dot on the curves represents a collocation point --- and since the flow is steadily decaying, a logarithmic profile is detected for a very brief time interval and a very small range only. Note that the establishment of this approximate logarithmic scaling is a transient behaviour. For later times (and thus a further decay of the shear mode) we could not detect such scaling. As a result, we obtain $\kappa_u$ and $C^+_u$ as 0.8 and 7.0 respectively, as annotated in the figure.

Furthermore, we rescale the temperature profiles shown in panel (c) using $u_\tau$ given in \eref{eq:wall_scale} along with the wall temperature unit, $T_\tau$, defined in terms of the wall heat flux as
\begin{equation}
T_\tau = \frac{Q}{u_\tau}, \quad\quad \mbox{where} \quad\quad
Q = \frac{Nu_w \kappa \Delta}{H} = -\kappa \frac{d\langle T \rangle_{A,t \in \mathcal{T}}}{dz}\biggr|_{z=0}.
\end{equation}
The above quantities yield the wall temperature unit $T^+ = (T_b - \langle T \rangle_{A,t \in \mathcal{T}}) / T_\tau$ where $T_b = 1$ is the temperature at the bottom wall, resulting in the corresponding log-law for temperature as~\cite{Kader:IJHMT1981}
\begin{equation}
T^+(z^+) = \frac{1}{\kappa_\theta} \log{z^+} + C^+_\theta\,.
\end{equation}
The resulting rescaled profiles are also plotted in figure~\ref{fig:loglaw}(d), with the diffusive sublayer and the logarithmic region marked by dashed black lines as in panel (b). Similar to the velocity profiles, the temperature profiles also have a diffusive sub-layer defined by $T^+ = Pr z^+$~\cite{Kader:IJHMT1981}. The log-region however, has $\kappa_\theta = 2.45$ and $C^+_\theta = 7.5$, which differ from the expected values in literature (to be discussed at greater length in \sref{sec:pgrad}). The situation is thus similar to the velocity case.

Finally, the time series of the wall-based Nusselt number, $Nu_w$, is reproduced in the last panel (e) with dashed vertical lines colored in correspondence with the four time instants, at which the profiles have been obtained. We observe that the heat flux is reduced when the velocity profiles indicate a turbulent shear-driven boundary layer. Since the heat flux is driven mainly by plumes, the stronger local shear disrupts this vertical transport of heat, resulting in a lower Nusselt number value. This reduction in turbulent heat transport by large-scale wind has also been observed in the case of rotating convection with free-slip plates by von Hardenberg et al.~\cite{Hardenberg:PRL2015}, where the time-series of $Nu$ showed multiple local minima corresponding to short durations when the horizontal winds became especially strong. Here, it is again a transient behaviour in the process of the decay of the shear mode. To further investigate the effects of shear flow, we consider the case of steady horizontal velocity forcing in the next subsection, which is applied to maintain the shear flow mode.

\subsection{Steady forcing by sinusoidal shear flow mode}
\label{sec:forced}

\begin{figure}
  \centerline{\includegraphics[width=1.0\textwidth]{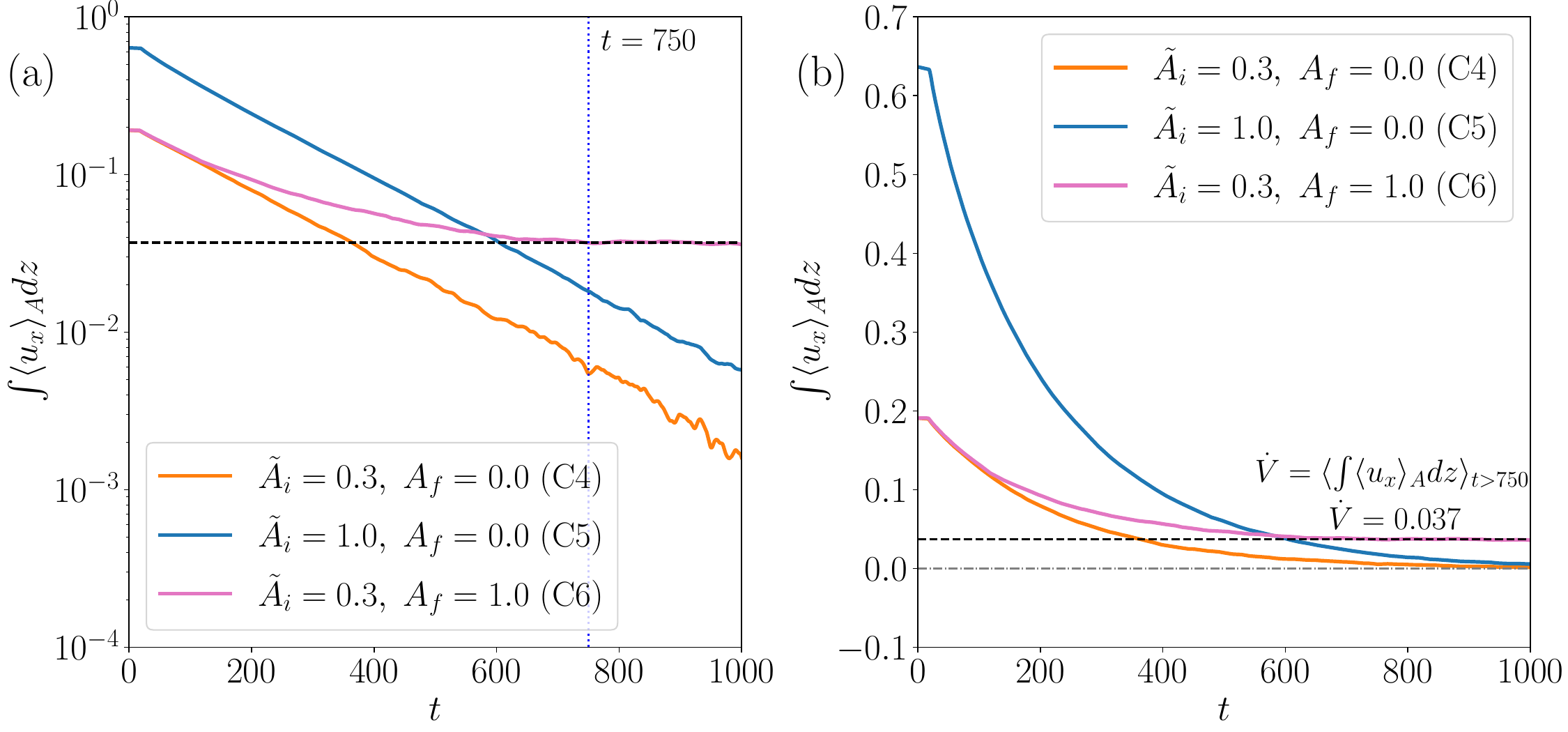}}
  \caption{Comparison of the decay of mean flow for the sinusoidally forced case along with   cases C4 and C5 shown previously in figure~\ref{fig:profile_decay}(f).
           The forced case attains steady mean flow at $t \approx 750$, and the mean shear flow rate at steady state is approximately 0.037 (marked with dashed black line).
           The decay of the flow is also shown in linear scale (b) to highlight the difference in mean flow rate between the forced and decaying cases.
          }
\label{fig:forced_decay}
\end{figure}

We now impose a continuous shear force along the positive $x$-direction, ${\bm f}=f(z){\bm e}_x$, to maintain a shear flow across the hot and cold plates. This additional term is introduced in the dimensional form of the momentum conservation equation, \eref{eq:u}, presented earlier in \sref{sec:numerics},
\begin{equation}
\frac{\partial {\bm u}}{\partial t} + ({\bm u} \cdot {\bm \nabla}) {\bm u} = -\frac{{\bm \nabla} p}{\rho} + \alpha g (T-T_{\rm ref}) \hat{{\bm z}} + \nu \nabla^2 {\bm u} + {\bm f}.
\label{eq:force_nse}
\end{equation}
The imposed body force is conservative, so that ${\bm f} = -\nu \nabla^2 {\bm U}_0$ \cite{Waleffe1997,Schumacher2001}. The total velocity field can therefore be decomposed into a mean flow with fluctuations, ${\bm u} = {\bm U}_0 + {\bm u}'$,
where
\begin{equation}
{\bm U}_0(z) = A_f \sin{\left(\frac{\pi z}{H}\right)} \hat{{\bm x}}, \qquad\qquad {\bm f} = -\nu \nabla^2 {\bm U}_0 = \frac{\nu \pi^2}{H^2} {\bm U}_0.
\end{equation}
We note that this steady shear has exactly the same form as the mode, which was added to the initial condition in cases C2 to C5 in the last subsection. In dimensionless form, we obtain the following form of the momentum equation,
\begin{equation}
\frac{\partial {\bm u}}{\partial t} + ({\bm u} \cdot {\bm \nabla}) {\bm u} = -{\bm \nabla}p + T \hat{{\bm z}} + \sqrt{\frac{Pr}{Ra}} \, \nabla^2 {\bm u} + A_f \pi^2 \sqrt{\frac{Pr}{Ra}} \sin{(\pi z)} \hat{{\bm x}}.
\end{equation}
We use the same flow setup as in the previous sections, with a Cartesian box of $\Gamma = 4$, at $Ra = 10^9$ and $Pr = 0.7$.
Even with the applied steady volume forcing, the initial sine flow decays for several hundreds of free-fall units before attaining a statistically steady state. The turbulent flow extracts kinetic energy out of the sustained shear mode, thus reducing its amplitude. This is further detailed in figure~\ref{fig:forced_decay}. As observed earlier, the time taken for attaining the statistically steady state increases with the size of the initial amplitude $\tilde{A}_i$. Furthermore, when the applied forcing amplitude $A_f$ is not sufficiently large, the steady-state mean flow rate is significantly smaller, which again leads to a long transient to reach the statistically steady state of mixed convection. Therefore, we choose a moderate $\tilde{A}_i=0.3$, similar to case C4, but a larger $A_f=1.0$ (similar to case C5), to set up the mixed convection run, C6. See again table \ref{tab:shear}. 

\begin{figure}
  \centerline{\includegraphics[width=\textwidth]{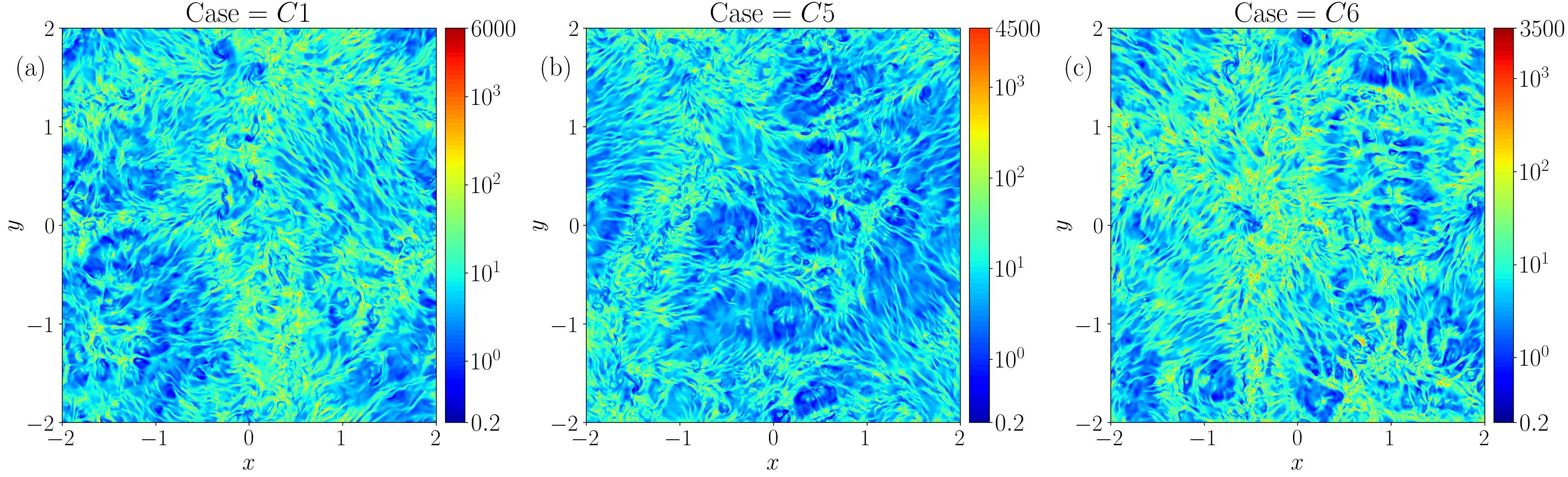}}
  \caption{Contour fields of the local Richardson number near the wall, $Ri_w$, as given by equation~\eref{eq:wall_Ri} for cases C1, C5 and C6.
  The snapshots shown here are at times $t=550$, 1000 and 1000, respectively. Each run is in his respective statistically steady state.}
\label{fig:wall_Ri}
\end{figure}

In figure~\ref{fig:forced_decay}, we compare the decay of the flow rate $\int_0^1 \langle u_x \rangle_A dz$ of C6 with that of runs C4 and C5, plotted earlier in figure~\ref{fig:profile_decay}(f). The forced convection case attains a steady mean flow at $t \approx 750$, and the time-averaged flow rate $\dot{V} = 1/(t_1-t_0)\int_{t_0}^{t_1} \int_0^1 \langle u_x \rangle_A dz dt \approx 0.037$, which is marked with a dashed black line. This is compared to $\int_0^1 U_{0,x}(z)dz\approx 0.19$. When viewed with linear scale in figure~\ref{fig:forced_decay}(b), the difference in mean flow between the forced and unforced cases is clear. However, despite this difference, $Re = 7440 \pm 120$, which differs in mean value by only 90 units from that of the base RBC case, C1, for which $Re = 7350 \pm 170$. The average values of $Nu$ and $Re$ for all the cases are listed in table~\ref{tab:shear} along with the averaging intervals used for each case. The table includes an additional case C7, which will be introduced in \sref{sec:pgrad}. Note that the averaging intervals are shorter for the decaying cases with high initial amplitude (cases C4 and C5), because the flow passes through a long transient to the statistically steady state. The table shows that the Nusselt numbers are almost identical for all cases, the Reynolds numbers overlap within the error bar. 

From the measured values of $Re$ in table~\ref{tab:shear}, one can also compute the Richardson number of the flow as
\begin{equation}
Ri = \frac{Ra}{Pr Re^2}\,.
\label{Rich}
\end{equation}
We obtain $Ri \approx 25$ for the present parameters, indicating that the convective flow is not dominated by the shear effects. The choice of a weak, but finite-amplitude sinusoidal flow here is deliberate, since our focus is on thermal convection and the possibility of obtaining an enhanced heat transport through the internal formation of a turbulent boundary layer, not imposed from the outside by a strong additional shear. In the cases where the mixed convection flow is dominated by a strong shear flow component, which would correspond to $Ri < 1$, it is possible to trigger an enhanced transport of heat from the walls as reported in \cite{Pirozzoli:JFM2017,Hamman2018,Schaefer:JFM2022}, and this will be explored in \sref{sec:pgrad}. However in natural convection, which is dominated by thermal plume formation and detachment from the walls,
such an intense transverse flow perpendicular to the growth direction of plumes is unlikely to form.

Furthermore, we can compute a local Richardson number near the wall, $Ri_w$ as
\begin{equation}
Ri_w(x,y) = \frac{\alpha g \Delta T \delta_T(x,y)}{2 u_\tau^2(x,y)},
\label{eq:wall_Ri}
\end{equation}
which can be defined in terms of the friction Reynolds number, $\widetilde{Re}_\tau$, and the near wall Rayleigh number, $Ra_w$, as
\begin{equation}
\widetilde{Re}_\tau(x,y) = \frac{u_\tau(x,y) \delta_T(x,y)}{\nu}, \qquad Ra_w(x,y) = \frac{\alpha g \Delta T \delta_T^3(x,y)}{2\nu\kappa},
\end{equation}
\begin{equation}
Ri_w(x,y) = \frac{Ra_w(x,y)}{Pr \widetilde{Re}_\tau^2(x,y)}.
\end{equation}
Both $u_\tau(x,y)$ and $\delta_T(x,y)$ are computed from the velocity and temperature gradients at the wall respectively,
employing a local formulation of eq. \eref{eq:wall_scale}.
\begin{equation}
u_\tau(x,y) = \sqrt{\sqrt{\frac{Pr}{Ra}}\, \frac{\partial u_h(x, y)}{\partial z}\Bigg|_{z=0}},\qquad
\delta_T(x,y) = \frac{1}{2}\Bigg|\frac{\partial T(x, y)}{\partial z}\Bigg|^{-1}_{z=0},
\end{equation}
where $u_h(x, y) = \sqrt{u_x^2(x, y) + u_y^2(x, y)}$, see also \cite{Scheel2014}.
A contour field of $Ri_w$ is plotted in figure~\ref{fig:wall_Ri} for cases C1, C5 and C6.
It is observed that the maximum $Ri_w$ falls steadily with increasing shear flow,
however the minimum $Ri_w$ is not sufficiently depressed by the mean flow.
Taking the area-averaged value of $u_\tau(x,y)$ at the bottom plate, we get $\bar{u}_\tau = \langle u_\tau(x,y) \rangle_{A,z=0} = 0.0283$ for the forced RBC case. Using the half-cell height $H/2$ as the length scale, we get the mean friction Reynolds number, $Re_\tau = \bar{u}_\tau H/ 2\nu \approx 535$. This magnitude should indicate that the flow in the boundary region is sufficiently turbulent.

\begin{figure}
  \centerline{\includegraphics[width=\textwidth]{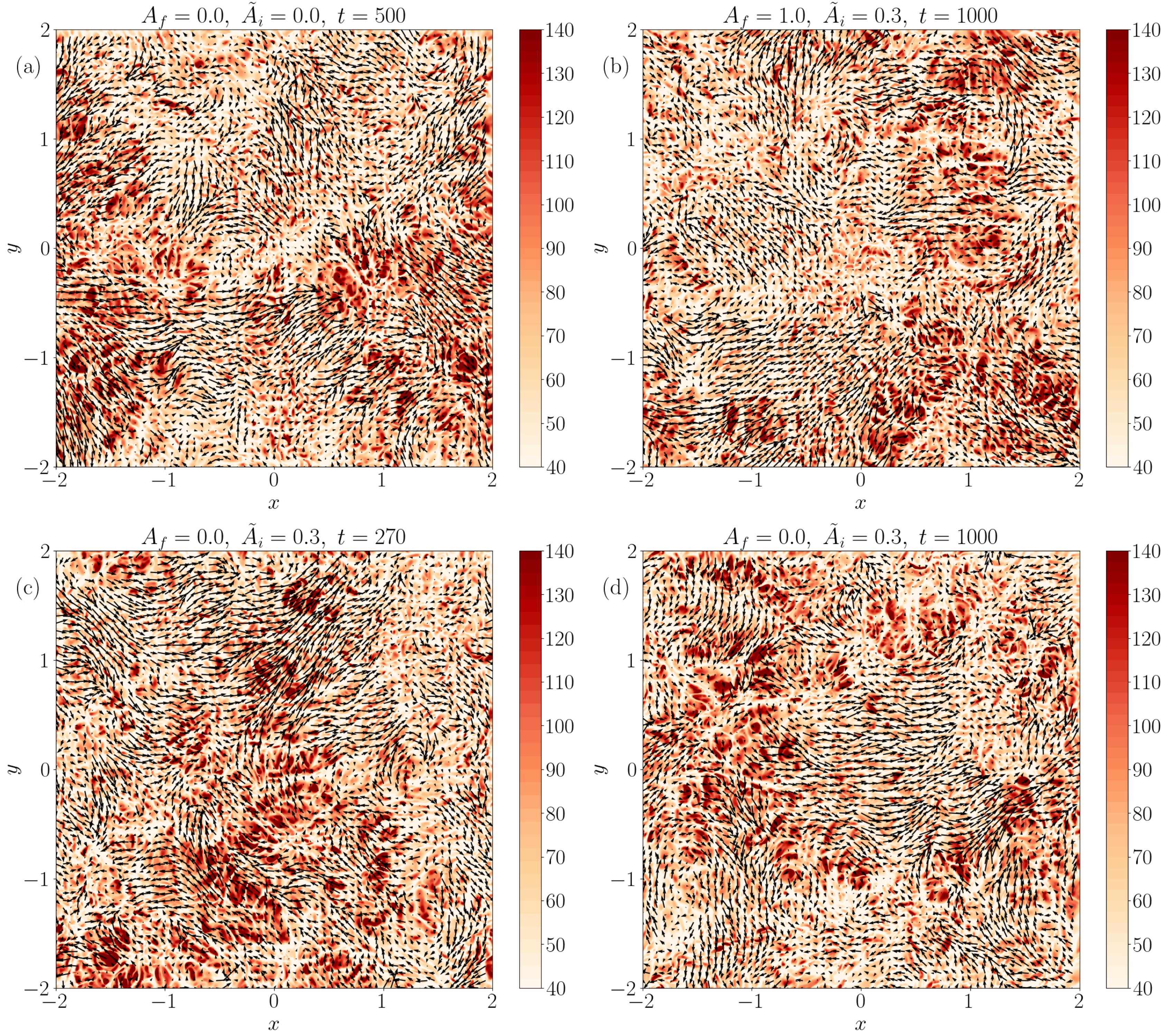}}
  \caption{Instantaneous velocity fields of cases C1, C4 and C6 at $z = 2\delta_T$ at selected times. The vector arrows project ${\bm u}$ into the horizontal $x$-$y$ plane. The background contours stand for $\partial T/\partial z$ at $z=0$. (a) The basic RBC case, C1. (b) The mixed convection case, C6 at the statistically steady state. (c,d) The decaying shear mode case, C4,  with an early stage of the flow at $t=270$ in panel (c), and towards the end of the simulation run at $t=1000$ in panel (d). }
\label{fig:forced_blflow}
\end{figure}

\begin{table}
  \begin{center}
    \begin{tabular}{|l|c|c|c|c|c|c|c|}
    \hline
    Case & $\tilde{A}_i$ & $A_f$ & $t_0$ & $t_1$ & $\Delta t$ & $Nu$ & $Re$ \\
    \hline
    C1 & 0.00 & 0.0 &  150 &  550 & 400 & $60.19 \pm 0.74$ & $~7350 \pm 170$ \\
    C2 & 0.01 & 0.0 &  250 &  550 & 300 & $60.42 \pm 0.83$ & $~7360 \pm 130$ \\
    C3 & 0.10 & 0.0 &  800 & 1000 & 200 & $60.66 \pm 0.77$ & $~7350 \pm 180$ \\
    C4 & 0.30 & 0.0 &  900 & 1000 & 100 & $60.14 \pm 0.75$ & $~7420 \pm 160$ \\
    C5 & 1.00 & 0.0 &  900 & 1000 & 100 & $60.42 \pm 0.73$ & $~7340 \pm 120$ \\
    C6 & 0.30 & 1.0 &  750 & 1000 & 250 & $60.29 \pm 0.67$ & $~7440 \pm 120$ \\
    C7 &  -   &  -  & 1300 & 1400 & 100 & $83.13 \pm 0.98$ & $74400 \pm 800$ \\
    \hline
    \end{tabular}
    \caption{Details of the decaying and forced convection cases discussed in \sref{sec:shear}. $\tilde{A}_i$ and $A_f$ are the amplitudes of the initial sinusoidal flow and corresponding volume forcing respectively. $t_0$ and $t_1$ indicate the start and end of the averaging intervals in free-fall times, yielding an averaging interval of $\Delta t=t_1-t_0$. Finally $Nu$ and $Re$ are the mean Nusselt and Reynolds numbers computed over $\Delta t$, along with their standard deviations.}
    \label{tab:shear}
  \end{center}
\end{table}

The fact that convection is not dominated by the shear mode is further corroborated by the horizontal velocity fields plotted in figure~\ref{fig:forced_blflow}, which shows the contours of the temperature gradient $\partial T/\partial z$ at the bottom wall (corresponding to the local $Nu$), along with the velocity field at $z = 2\delta_T$ marked by arrows. Panels (a) and (b) show the steady-state flows of cases C1 and C6, respectively. Similarly to what we observed in figure~\ref{fig:area_frac}, the flow patterns are not very dissimilar between the two cases, indicating that convection still dominates the dynamics in the boundary regions, especially with the modestly high $Ra$ and volume forcing, which just maintains the mean flow without completely negating the effects of buoyancy. Near the top and bottom walls, this weak shear flow approximates the effect of the `mean wind', which is observed in confined convection with large-scale circulation. Panels (c) and (d) show the decaying case C5 at an early stage when there exists a strong mean flow (c), and at the end of the simulation (d), when the mean flow has significantly diminished. We surmise that even with a strong large-scale circulation that can be expected at high $Ra$, the boundary layers continue to retain characteristics of plane-layer RBC. Consequently, despite the high $Ra$ and the augmented mean flow, the heat transfer is not markedly affected. See again table \ref{tab:shear}.

\begin{figure}
  \centerline{\includegraphics[width=0.8\textwidth]{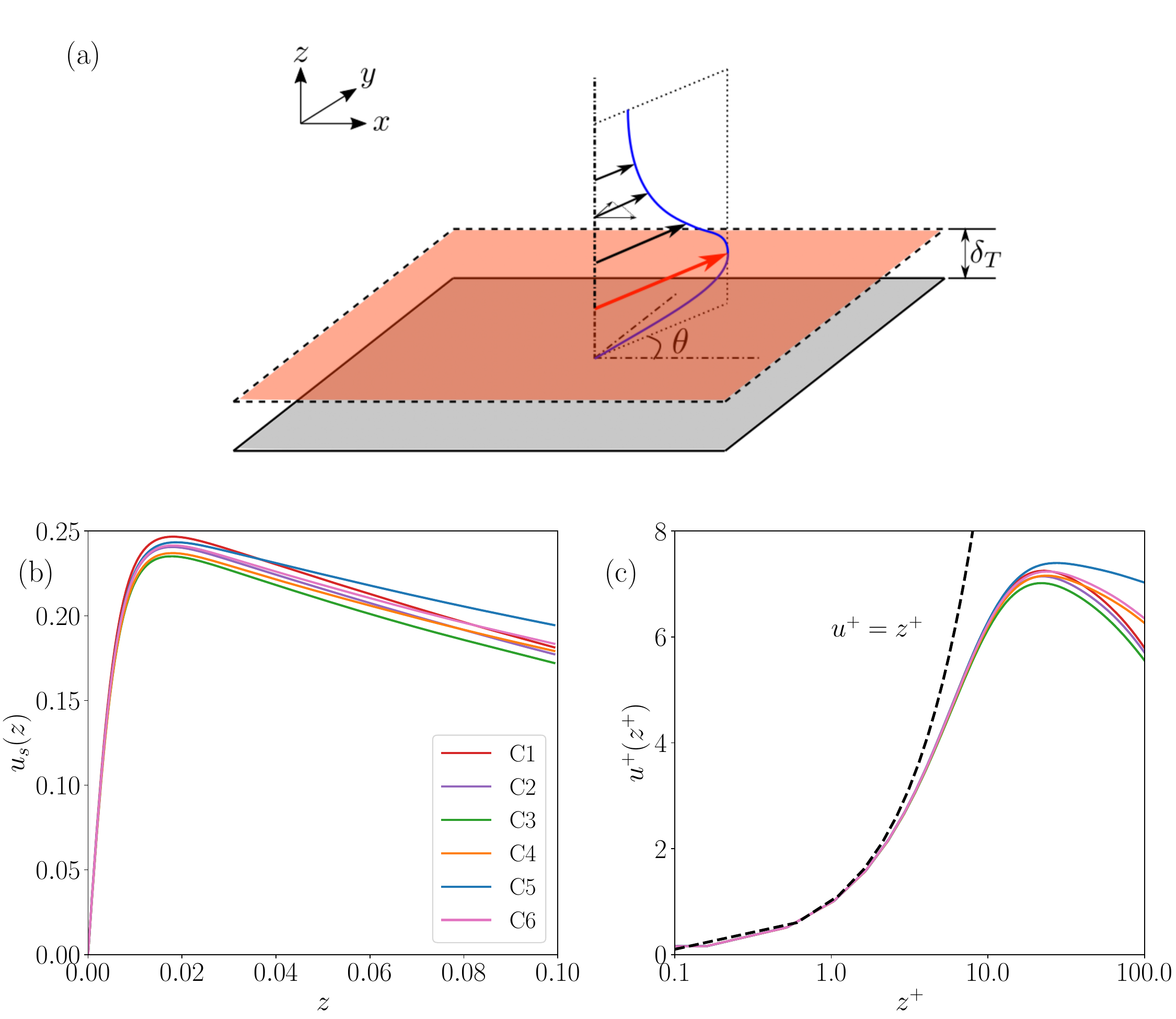}}
  \caption{Conditional mean velocity profile analysis. We plot the mean velocity in wall-units for shear-dominated regions. (a) Principal sketch of the analysis. The direction is determined at the red colored plane at $z=\delta_T$. The profile is computed pointwise by projecting the velocity onto a vertical plane aligned with the direction of the flow at the thermal boundary layer - the angle of this plane is therefore the same as contours plotted in figures~\ref{fig:uxField} (top row) and ~\ref{fig:area_frac}. (b) Unscaled mean profiles for the long-term state. (c) Replot of the same data in wall units. A log-profile is absent. Instead the velocity field in the shear-dominated regions are similar to that of a wall-jet.}
\label{fig:localLog}
\end{figure}

Considering that the shear flow in the boundary region is a patchwork of differently oriented local shear flow regions, we can probe the existence of a log-layer by projecting the velocity field onto a plane aligned with the direction of the flow at the edge of the thermal boundary layer. This is sketched in figure~\ref{fig:localLog}(a). A similar analysis of the velocity profile along a dynamically orienting plane aligned with the large-scale circulation (LSC) was performed for a cylindrical cell in \cite{Scheel2017}. Such an analysis is partly motivated by the fact that in flat-plate boundary layers, the logarithmic velocity profile is obtained for the velocity aligned in the streamwise direction of the flow.

The resultant velocity profiles, conditioned to the coherent regions of the near-wall layer, are shown in figure~\ref{fig:localLog}(b) for all cases from C1 through C6. The corresponding profiles rescaled in wall-units, eq.~\eref{eq:wall_scale}, are plotted in panel (c). The profiles obtained are similar to those of a turbulent wall-jet rather than a turbulent flat-plate boundary layer. This is consistent with the earlier observations for a plane aligned along the LSC roll~\cite{Scheel2017}.

\subsection{Steady forcing by constant pressure gradient}
\label{sec:pgrad}

\begin{figure}
  \centerline{\includegraphics[width=0.99\textwidth]{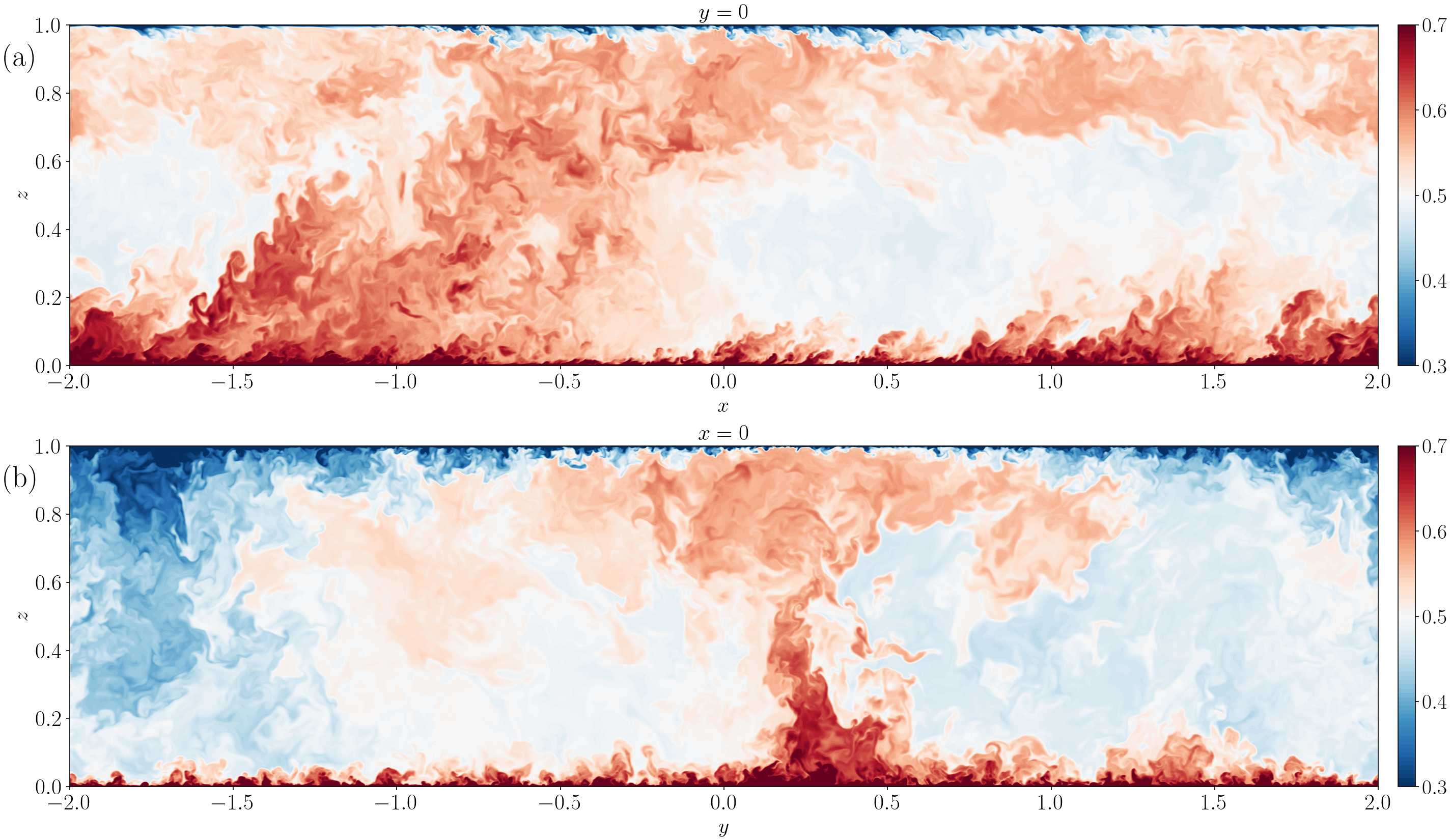}}
  \caption{Temperature contours for case C7. (a) Streamwise-wall normal cut at spanwise midplane. (b) Spanwise-wall normal cut at streamwise midplane. The shear flow sweeps the plume (or plume cluster) strongly inclined across the plane in panel (a). The spanwise cross-section in panel (b) indicates that the plumes are constrained to rise from a narrow strip along the centerline of the bottom plate between the counter-rotating streamwise rolls.}
\label{fig:case_C7}
\end{figure}

The log-profile for turbulent shear flow boundary layer is usually observed under a strong pressure gradient. We therefore subject our plane-layer RBC case to a mean pressure gradient along the $x$-direction, and this case is denoted C7 here. The mean pressure gradient is therefore imposed in equation~\eref{eq:force_nse} as $f_x = |\partial p_0 / \partial x| = 0.02$ in dimensionless units. The initial condition for this case is the statistically steady state solution of C6. The resultant shear-flow sweeps the thermal plumes across the plate, and overwhelms the effects of buoyancy as shown in figure~\ref{fig:case_C7}(a). Figure \ref{fig:case_C7}(b) shows a spanwise vertical cut and identifies two counter-rotating circulations, which seem to point into the streamwise direction. The plume dominated region of the plate is restricted to a narrow region along the centerline of the plate.

\begin{figure}
  \centerline{\includegraphics[width=0.9\textwidth]{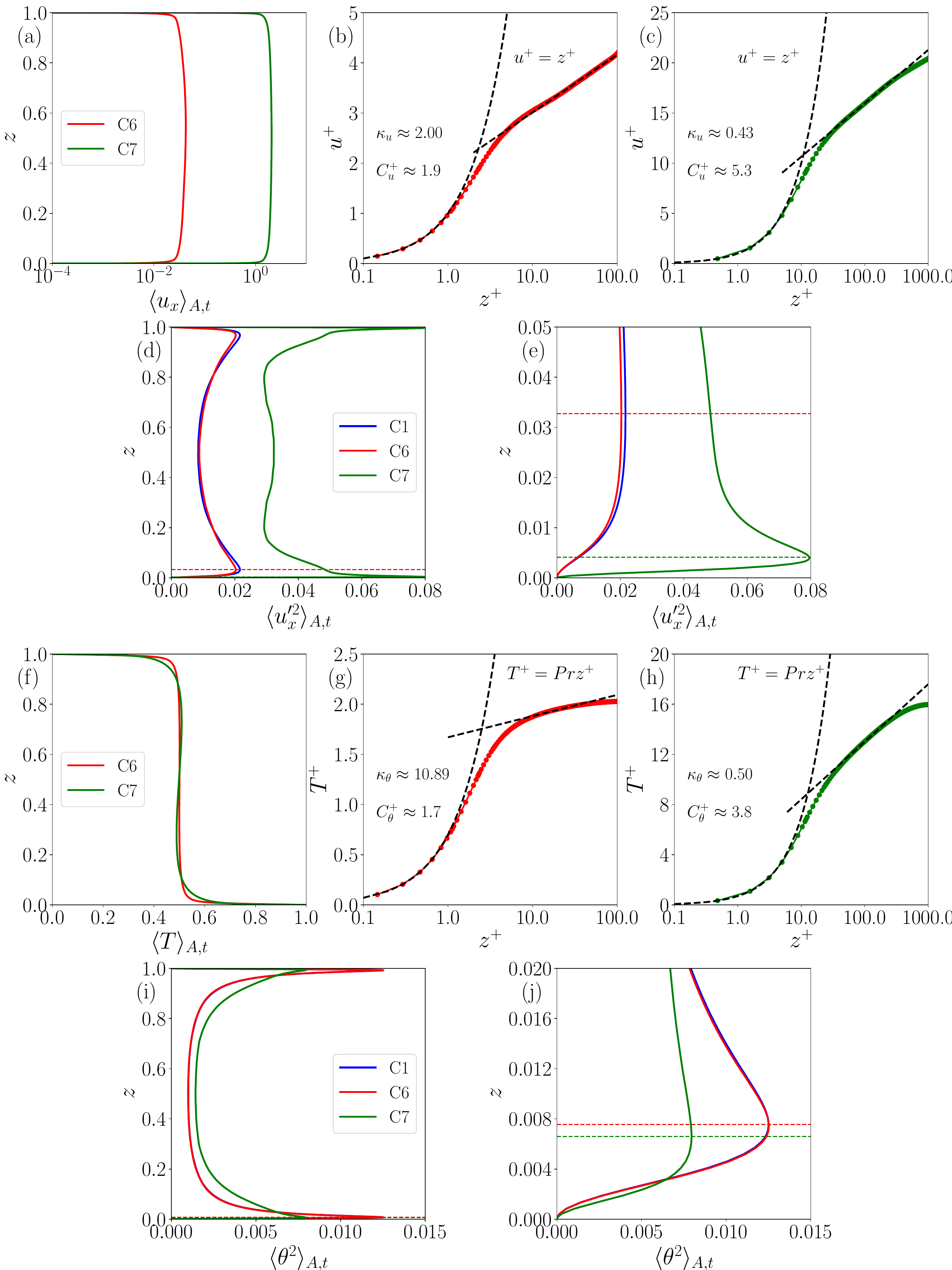}}
  \caption{Comparison of velocity and temperature mean profiles of the unforced and steady forced cases. (a,f) Mean profiles of C6 and C7 across whole layer. (b,g) Mean profiles replotted in wall units for C6. (c,h) Corresponding profiles replotted in wall units for C7. The streamwise velocity fluctuation profiles, $\langle u_x^{\prime\,2} \rangle_{A,t}$, across the whole layer and near the bottom wall are plotted in (d) and (e) respectively for cases C1, C6 and C7. The corresponding temperature fluctuation profiles, $\langle \theta^2 \rangle_{A,t}$, are shown in (i) and (j) respectively. The resulting fluctuation thickness values are marked with dashed lines.}
\label{fig:forced_loglaw}
\end{figure}

Computing the shear-dominated and plume-dominated regions of the boundary region for this case, we get almost an inversion of the area-fraction observed earlier. Whereas the cases from C1 through C6 had an approximately 60\% plume dominated area, case C7 has only 45\% incoherent region, and 55\% shear dominated region. The wall Richardson numbers calculated from~\eref{eq:wall_Ri} also shows a strongly shear dominated boundary layer -- the maximum $Ri_w$ on the plate is down by 2 orders of magnitude to less than 100. Also $Ri\approx 0.25$ now, cf. eq. \eref{Rich} and table \ref{tab:shear}. It is important to stress that we are far from the regime of thermal convection with case C7. Instead, we have a forced channel flow with heat-transfer -- there are no well defined plume ejection regions anymore. The scope for the formation of turbulent superstructures~\cite{Pandey2018} is also greatly diminished owing to the sweeping flow which disrupts such coherent structures whose boundaries are aligned perpendicular to the plates.

In figure~\ref{fig:forced_loglaw}, we compare the velocity ($x$-component) and temperature profiles of cases C1, C6 and C7. The profiles for C6 are obtained from its statistically steady state regime, for times $t > 750$, see also figure~\ref{fig:forced_decay}. The velocity profiles shown in figure~\ref{fig:forced_loglaw}(a) are plotted with log-scale to reflect the enormous difference in the strength of forcing between C6 and C7. Furthermore, case C7 shows the flat profile of turbulent Poiseuille flow, typically obtained by a uniform pressure gradient~\cite{Moser:PoF1999}. The near-wall velocity profiles in wall-units, see eqns.~\eref{eq:wall_scale} and~\eref{eq:wall_logu}, for C6 and C7 are plotted in panels (b) and (c) respectively, since both profiles have starkly different ranges. Their corresponding coefficients $C^+_u$ and $\kappa_u$ are calculated using least-square fits and annotated within the respective panels. Note that for C7, the coefficients $C^+_u = 5.3$ and $\kappa_u = 0.43$ match reasonably well with the values for turbulent channel flows~\cite{Schlichting:book:BL}. The streamwise velocity fluctuation profiles are plotted in figures~\ref{fig:forced_loglaw}(d) and (e) for the full height of the layer and close to the bottom plate respectively. The velocity boundary layer, marked here by the height of maximum velocity fluctuations from the wall, is much thinner at approximately $0.0041H$. Nevertheless, we have 12 collocation points across this height, indicating that the thin boundary layer is reasonably well-resolved.

The mean temperature profiles in figure~\ref{fig:forced_loglaw}(f) show a smaller difference between cases C6 and C7. However, the flat well-mixed profile in the bulk typically observed in RBC is slightly inclined by the shear flow. The mean temperature falls slightly below 0.5 in the bottom half of the plane, and slightly above 0.5 in the top half. The logarithmic layers of the mean temperature in wall units are plotted in panels (g) and (h). While case C6 has only a weakly discernible log-layer, case C7 shows a distinct corresponding profile. It has been empirically observed from experimental data by Kader~\cite{Kader:IJHMT1981} that $C^+_\theta$ is a function of $Pr$
\begin{equation}
C^+_\theta = (3.85 Pr^{1/3} - 1.3)^2 + 2.12 \ln{Pr}.
\end{equation}
For $Pr=0.7$, we get $C^+_\theta = 3.73$, which agrees very well with $C^+_\theta = 3.8$ for C7 and annotated in panel (h). Similarly, the obtained value of $\kappa_\theta = 0.5$ also matches well with the expected value of 0.47 from literature~\cite{Kader:IJHMT1981}.

The temperature fluctuations plotted in figure~\ref{fig:forced_loglaw}(i) and (j) show that the profiles and fluctuation thickness for C1 and C6 collapse together, whereas C7 has a slightly reduced maximum fluctuation amplitude due to the strong shear and comparable boundary layer thickness. Interestingly, the thermal BL thickness is not increased in this mixed convection case. Overall, the strong pressure gradient and the resulting strong shear distinctly alters the boundary layer structure of convection. Furthermore, we also note from table~\ref{tab:shear}, that the $Re$ for C7 is an order of magnitude higher than the base RBC case. This increased $Re$ also leads to an enhanced mixing of heat in the bulk (which manifests in the mean temperature falling below 0.5 near the bottom plate and above 0.5 near the top plate as observed above). Consequently, we observe an elevated value of $Nu$ for this case. Similar experiments at lower $Ra$ have also observed this increase, specifically when the convective flow is completely overtaken by the strongly forced transverse shear flow~\cite{Pirozzoli:JFM2017}.

\begin{figure}
  \centerline{\includegraphics[width=0.85\textwidth]{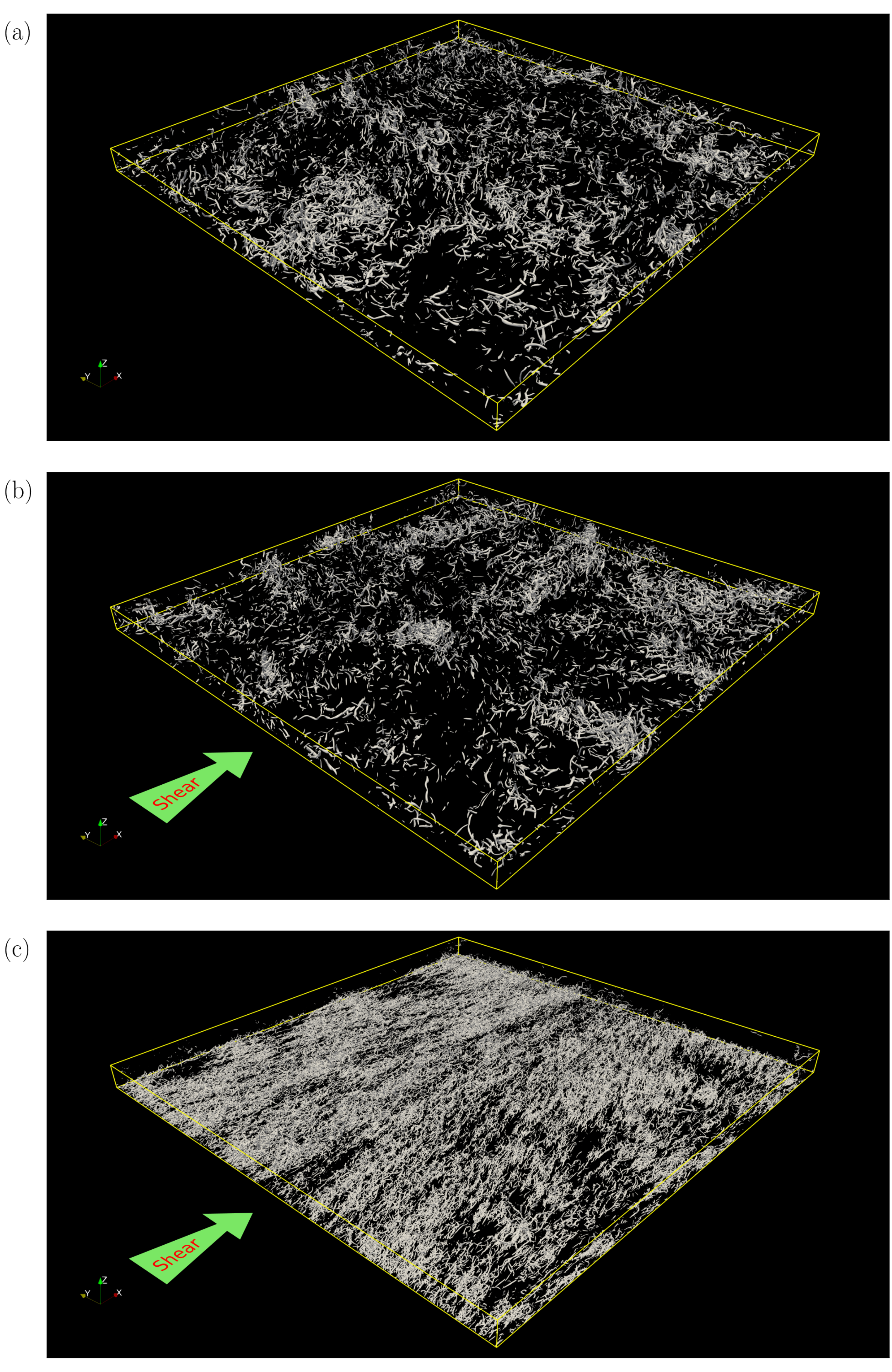}}
  \caption{Iso-surfaces of the second invariant of the velocity gradient tensor $Q$, also known as the $Q$-criterion, plotted for a thin section of the flow near the bottom wall over a subdomain of size $4 \times 4 \times 0.2$ in units of $H$ for cases C1 (a), C6 (b) and C7 (c). The isosurfaces correspond to $Q=125$ in (a) and (b), whereas $Q=1000$ for (c). The shear flow direction is indicated by the green arrows.}
\label{fig:compare_q}
\end{figure}

Finally, we show the iso-surfaces of $Q=(\|\bm \Omega\|_2^2-\|\bm S\|_2^2)/2$ with $\bm \Omega $ and $\bm S$ the anti-symmetric and symmetric parts of the velocity gradient tensor for cases C1, C6 and C7 in figure~\ref{fig:compare_q}(a), (b) and (c), respectively. The iso-surfaces are plotted for a thin slice of the domain close to the bottom plate, of size $4 \times 4 \times 0.2$ in units of $H$. For cases C1 and C6, the iso-surfaces were plotted for $Q=125$, whereas for C7, choosing such a low value of $Q$ results in a dense network of streamwise and hairpin vortices, completely obscuring the plate. So panel (c) has its iso-surfaces plotted at $Q=1000$ instead. The direction of shear flow for cases C6 and C7 are indicated by the green arrows.
While the vortex tubes are mostly aligned vertically for C1, the mean flow sweeps the structures to the right for the shear-flow cases, weakly only for case C6, but very pronounced for C7.
This impact of shear flow on plume structures, which causes them to align with the flow and also affects their mean plume-spacing has also been studied in particle image velocimetry experiments~\cite{Shevkar:PRF2019}.

\section{Conclusions and outlook}
\label{sec:conclusions}

We presented results on the turbulent near-wall fluctuations and the sensitivity to initial conditions based on three-dimensional direct numerical simulations of plane-layer RBC for a range of Rayleigh numbers $Ra$ that span 6 orders of magnitude. We quantified the effect of an additional streamwise forcing on heat and momentum transfer and mean profiles. By focusing on the temperature and velocity fluctuation profiles and their scaling with respect to $Ra$, we observe a transition in the scaling laws at around $Ra \sim 10^7$, which seems to coincide with the transition to the so-called hard turbulence regime of RBC~\cite{Castaing:JFM1989,Emran:JFM2008}. Interestingly, the scaling exponents of the fluctuations at different distances from the wall do practically not differ in the near-wall region. Only temperature fluctuations drop stronger with increasing $Ra$ when moving to the bulk, caused by the enhanced turbulent mixing. This suggests once more that the dynamics is primarily driven by the differences in fluid flow and organization in the near-wall regions and complements our previous studies on this flow geometry in \cite{Samuel2024} and \cite{Shevkar2025}. 

The near-wall region is decomposed into local shear-dominated coherent and plume-dominated incoherent regions. Their area fraction is determined in three different ways, two based on the velocity and one based on temperature. Thus, we can also connect temperature and velocity field analysis. In all cases, the numbers for the area fraction agree to the same order of magnitude. More importantly, all measures detect an approximate independence of the Rayleigh number, which seems to be connected to the self-similarity of the thermal plume network as found in \cite{Shevkar2025}. The mean spacing of line-like thermal plumes, the essential building block of BL, and thermal boundary layer thickness $\delta_T$ appear in a nearly constant ratio with respect to each other. We add that this ratio is bounded from below by the relative critical wavelength (measured in units of $\delta_T$) of a linear stability analysis, as shown in \cite{Shevkar2025}. Our present findings thus underline that the near-wall region in RBC is organized in a bottom-up process.

Expanding on the work done in~\cite{Samuel2024}, we also explore the scaling of $Nu$ in the coherent (shear-dominated) and incoherent (plume-dominated) regions of the BL flow. A significant observation to be made here is that the contribution of the plume-dominated regions to overall heat transfer increases with an increase in $Ra$. The present results confirm our previous findings of fluctuation-dominated near-wall layers in the plane-layer RBC configuration. Enhanced heat transfer in plume-dominated regions was also found in 2D studies of RBC in \cite{Zhu:PRL2018}. Here, the result is extended to the three-dimensional case.

Furthermore, we present results on the effect of decaying shear flow mode on the distribution of plumes in the boundary region by tracking the area fractions of coherent and incoherent regions. We see that due to the wave-like organization of the turbulent flow, the area fraction is not significantly altered by the additional shear mode. Consequently, the heat and momentum transfer rates remain unaffected by this additional initial flow mode of finite amplitude. We also notice that the decaying flow mode induces short-term transient signatures of a logarithmic profile (with non-standard fit parameters) in the early stages of the dynamical evolution.

All the different {\em finite-amplitude} sinusoidal shear mode-type perturbations of the initial diffusive equilibrium of the plane layer decayed with time. The turbulent RBC state ends up in the same turbulent attractor for our DNS series. We analyse this finding by the global heat and momentum transfer, i.e. $Nu$ and $Re$, both of which remain the same within the error bars as documented in table \ref{tab:shear}. We thus did not find footprints of different coexisting macrostates of RBC. One could criticize that our Rayleigh number of $Ra=10^9$ is still too small for a detection of a bistable regime of turbulent convection. However, it should be recalled that in the mentioned analogy to the transition to turbulence in wall-bounded flows, such precursors of the bistable nature with hysteretic transitions are detectable at Reynolds numbers much lower than the actual critical Reynolds number (determined in a statistical way), see e.g. \cite{Avila2023}. Note also that the near-wall layers in RBC fluctuate strongly for the whole range of Rayleigh numbers, both in shear-dominated regions as well as plume-dominated ones. This leaves to our view the question on the actual mechanisms that would trigger such a hysteretic transition open.

To continue along these lines, we maintained the initial shear mode with sinusoidal volume forcing that is added to the momentum balance of the Boussinesq model. A unidirectional mean flow is then generated with a logarithmic profile (again, with non-standard coefficients) for mean velocity and temperature. Interestingly, the existence of these mean flow profiles do not affect the global heat transfer which is measured by $Nu$. The momentum transfer is slightly enhanced, but the Reynolds numbers still overlap within the error bars. We thus found for DNS case C6, two logarithmic near-wall layers without impact on the global transport of heat and momentum. We conclude that this volume forcing, although containing a finite perturbation amplitude, seems to be too weak (also indicated by the actual values of von K\'arm\'an and offset constant); buoyancy effects remain dominant as indicated by the Richardson number analysis in the spirit of Pirozzoli et al. \cite{Pirozzoli:JFM2017}.

Indeed, a stronger shear forcing by a constant pressure gradient causes detectable differences bringing the system to $Ri<1$ with a fully established turbulent boundary layer. The BL displays a logarithmic scaling with close-to-standard fit constants for velocity and temperature. In this case, global heat and momentum transport are enhanced significantly. However, this takes the system far away from the realm of a natural convection flow into a forced convection regime. All these findings lead us to the conclusion that the near-wall layer RBC, which we investigated here in its simplest plane-layer configuration, is a non-standard boundary layer that still needs to be better understood, in particular for higher Rayleigh numbers, which might start at $Ra\gtrsim 10^{10}$.

\section*{Acknowledgments}
The work of R.J.S. and J.S. is funded by the European Union (ERC, MesoComp, 101052786). Views and opinions expressed are however those of the authors only and do not necessarily reflect those of the European Union or the European Research Council. Neither the European Union nor the granting authority can be held responsible for them. The authors also gratefully acknowledge the Gauss Center for Supercomputing e.V. (https://www.gauss-centre.eu) within the Large Scale Project Nonbou for funding this project by providing computing time on the GCS Supercomputer JUWELS at the J\"ulich Supercomputing Center (https://www.fz-juelich.de/en/ias/jsc). Special thanks to Mathis Bode at the J\"ulich Supercomputing Center, Forschungszentrum J\"ulich, for enormous help and assistance with computing resources. 
We thank Marc Avila, Erik Lindborg, Philipp Schlatter, Katepalli R. Sreenivasan, and Mahendra K. Verma for insightful discussions.
The data slices and analysis scripts which were used for the present work are available on Zenodo~\cite{Samuel:Zenodo2025}.

\section*{References}
\bibliographystyle{unsrt}

\end{document}